\documentclass[english,aps,manuscript,floatfix]{revtex4}
\usepackage[T1]{fontenc}
\makeatletter

\usepackage{float}
\usepackage{float}
\usepackage{float}
\usepackage{graphicx}
\usepackage{babel}
\makeatother

\begin{document}

\title{Simulation Study of Sulfonate Cluster Swelling in Ionomers. }

\author{Elshad Allahyarov}
\affiliation{Department of Physics, Case Western Reserve University, Cleveland,
44106 Ohio, USA\\
OIVTRAN, Joint Laboratory of Soft Matter, Moscow, 127412 Russia,\\ 
HHU D\"usseldorf, Institut f\"ur Theoretische Physik II,
  Universit\"atstrasse 1, 40225 D\"usseldorf, Germany } 
\author{Philip L. Taylor}
\affiliation{Department of Physics, Case Western Reserve University, Cleveland,
Ohio 44106, USA}
\author{Hartmut L\"owen}
\affiliation{ HHU D\"usseldorf, Institut f\"ur Theoretische Physik II,
  Universit\"atstrasse 1, 40225 D\"usseldorf, Germany }

\begin{abstract}

We have performed simulations to study how increasing humidity affects
the structure of  Nafion-like ionomers under conditions of low
sulfonate concentration and low humidity.   At the onset of membrane
hydration, the clusters split into smaller parts.  These  subsequently
swell, but then maintain constant the number of sulfonates per
cluster.   We find that the distribution of water in low-sulfonate
membranes depends strongly  on the sulfonate concentration. For a
relatively low sulfonate concentration, nearly all the side-chain
terminal groups are within cluster formations, 
and the average water loading per cluster matches the water content
of membrane.  However, for a relatively higher sulfonate concentration
the water-to-sulfonate ratio becomes non-uniform.  The clusters become
wetter, while the inter-cluster bridges become drier.  We note the
formation of unusual shells of water-rich material that surround the
sulfonate clusters.

\end{abstract}
\maketitle

\section{introduction}

An interest in ionomers, i.e. ion-containing polymers, began fifty
 years ago with the development of organic ion-exchange resins  
\cite{first-ionomer}. The properties of these materials are completely
 different from those of other polymers as a consequence of the
 ionization of the ionic groups in polar solvents. The specificity of
 the interaction between the ion, the solvent, and the polymer 
makes it possible for these materials to be used as perm-selective
membranes, thermoplastics or films for micro-encapsulation 
and coating \cite{perry}. A significant interest in ionomer materials
also stems from their growing application as a polymer electrolyte membrane
(PEM) or a proton exchange membrane in fuel cell technology
 \cite{sacca-2006,saito-2004,kreuer-2001,smitha-2005}.

DuPont de Nemours was the first manufacturer in the early 1960s to
develop a perfluorosulfonic membrane commercially
\cite{first-nafion,banerjee}. This membrane, which was named
Nafion$^{\tiny\textregistered}$, consists of a polytetrafluoroethylene
hydrophobic backbone to which perfluorovinyl ether pendant side chains
are attached at more or less equally spaced intervals. The 
pendant chains are terminated by sulfonic head groups SO$_{3}$H, and
these are responsible for the  large variety of microstructures in
which the ionomer can be assembled.  
When exposed to humidity, the membrane takes up large
amounts of water, leading to  the dissociation of the acid groups
SO$_{3}$H$\rightarrow$SO$_{3}^{-}$ + H$^{+}$ and to the formation of 
a nanophase-separated network
of aqueous (hydrophilic) clusters and hydrophobic polymer. 
According to the cluster morphology model of 
 Hsu and Gierke \cite{Gierke2}, spherical clusters are uniformly distributed
throughout the material and are  interconnected by channels
\cite{robertson,jalani2006}.
 Subsequent cluster-based models, such as the Mauritz-Hopfinger
model \cite{MOmodel,mauritz2004}, the Yeager three-phase model
\cite{yeager}, the Eisenberg model of clusters of hydrocarbon ionomers
\cite{eisenberg}, and the Litt model  of a lamellar morphology 
for sulfonate domains \cite{litt1997}, have  tried to quantify the cluster
radius and spacing as a function of the polymer equivalent weight and
the hydration level. Other structural models were proposed to describe
the membrane swelling process from a dry state to a colloidal
suspension as a continuous process \cite{Gebel,rohr}.

There is still ongoing debate about which one of the proposed models
is more suitable and effective in representing  the ionomer's
conductivity through its nanophase separated network of hydrophilic
regions. The issue is complicated by the fact that experimental
studies show the ionomer structure to depend on 
the pretreatment methods used in  its preparation
\cite{paddison-1998-epsilon,slade,banerjee,cappadonia-1994,efield-exp,epsilon-nafion}. 
The membrane pretreatment serves to reduce the remnant anisotropy in
the morphology of extruded membranes, and to clean a solvent-cast
membrane from impurities \cite{arcella-2005,mauritz2004}. 
Rigorously speaking, the question of how the 
pretreatment steps, such as swelling and/or  boiling in
solvents, annealing, rinsing in water, drying in vacuum/air, and
the order of these steps, affect the membrane
morphology, is not yet answered.  
Most of the pretreatment protocols have the ultimate goal of
improving the water uptake of the membrane \cite{yamamoto-2006,slade}. For
example, in Ref.~\cite{slade} it has been shown that the water uptake 
of a dry membrane depends on  how it was dried from its swollen
state at elevated temperatures. If it was first cooled and then dried,
then the membrane keeps its swollen volume. But if  it was first dried
and then cooled, the membrane shrinks in volume during the drying 
process. As a result, the outcome of the first protocol is a membrane
that takes up a desirably large amount of water, and thus has a better
proton conductivity.

 The water solvation of a PEM, which is necessary for its efficient
 operation, reduces its working temperature range: the membrane will
 not be exploitable at freezing and boiling water temperatures. A
 possible way  to overcome this limitation is the development of  new
 membranes that can operate at the low wetting conditions where
 $\lambda$, which is the number of water molecules per sulfonate
 group, is less than five.  In the case of full hydration 
there are 5  water molecules in the primary hydration shell of a
sulfonate \cite{fimrite2005,laporta-1999,lue-2009}.   
 In low-humidity membranes, the protons diffuse along narrow pathways
 near the SO$_{3}^{-}$ terminals of 
side chains, and two conflicting effects come into play. On the one
 hand, the proximity of negatively charged sulfonates 
considerably suppresses the mobility of the protons. On the other
hand, when the separation distance between sulfonates is small, the
activation energy for proton hopping between adjacent 
end-groups becomes comparable with the activation energy in the
bulk water \cite{spohr2006}, making the net result unclear. 
The proton mobility in low-humidity membranes can be also elevated   
by adding flexibility  to the sidechains, and by modifying the network
 structure of sidechain clusters. 

Despite the fact that various models have emerged to explain the properties
of hydrated Nafion membranes, a systematic study of how the molar
concentration $\eta$
of sulfonate head groups and the solvent content parameter $\lambda$
affect the network structure of sulfonates, and particularly the
swelling of single clusters has not yet appeared. This absence
 is important for understanding proton transport and the
onset of percolation  in low-humidity membranes, and forms the
motivation for this study.  

Here we perform simulations to investigate the dependence of the
cluster swelling on the hydration level $\lambda$ and the sulfonate
molar concentration $\eta$ of the membrane by  employing different
sidechain architecture models. 
We restrict ourselves to the case of ionomers for which there is   
 no bulk water inside the sulfonate clusters, and in which the sulfonate
concentration is considerably  below the percolation limit for the head
groups. It is expected that in these {\it low-humidity} and {\it low-sulfonate}
membranes  no overlapping between sulfonate 
clusters takes place. In order to distinguish a sulfonate cluster from a water
 cluster, which is necessary in the interpretation of our simulation
 results,  for the former we adopt the term ``sulfonate multiplet'',
 first introduced  by Eisenberg in Ref.~\cite{eisenberg} to describe
 the primary  aggregates of sulfonates. 
 We show that at the onset of swelling, which is 
defined as the transition from a dry multiplet into a wet multiplet with
dissociated protons, the multiplets split into smaller parts.
The solvation of these resultant multiplets is analyzed 
for different hydration levels and sulfonate concentrations. 
In particular, we will demonstrate the formation of water shells
around the sulfonate multiplets.

The paper is organized as follows. In Section II we briefly discuss
the benefits of using coarse-grained models (as opposed to all-atomistic
approaches), and describe the coarse-grained model and system
parameters employed here. The simulation details are outlined in
Section III.  Results on multiplet formation in dry and solvated
 membranes, water shells around multiplets, ionomer
deformation, and proton diffusion are discussed in Section IV. We
conclude in Section V.

\section{Coarse-grained system parameters}
\label{sys-parameters}
Despite the rich variety of experimental findings and theoretical
predictions for the ordered morphology in PEM
materials, numerical experiments have so far had little success in
finding any clear picture of cluster formation in hydrated
membranes. The  main reason for this is the fact that  individual
ionic clusters are about 2--5 nm in size, and this is usually
comparable to, or even larger than, the system 
sizes affordable  in all-atomistic modeling.   As a result,
atomistic simulations, which are quite helpful for understanding the
simple pore physics and small ionomer molecular conformations, are not
able to capture the distribution
of sulfonate clusters in the hydrophobic matrix. However, as already
outlined in the introduction, a knowledge of this
distribution is crucial for the determination
of the ionomer connectivity and the proton conductivity of the PEM
material. 

Fortunately polymers show a large degree of universality in their
static and dynamic behavior. The universal scaling properties
of the ionomer as a function of  chain length, sulfonate
density, and membrane composition can be most efficiently studied via
coarse-grained molecular models. 
One of the most commonly employed systems is a bead-spring model,
where each bead  represents a segment of a realistic chain.
Wescott et al \cite{wescott2006} and Vishnyakov et al
\cite{vishnyakovDPD} have simulated large ionomer systems using
coarse-grained approaches in which an entire sidechain was represented
by a nanometer-size hydrophilic blob. Their simulations report
irregularly shaped hydrophilic clusters embedded into the polymeric
matrix of backbone chains. While such gross coarse-graining is
computationally convenient, it is not possible to draw firm
conclusions regarding proton diffusion from the conformational results
obtained for the polymer.  It is therefore necessary to limit the 
coarse-graining approach to the level at which the sulfonic acid
groups  of the polymer can be explicitly treated, as these groups
contain the  essential membrane-specific interaction sites relevant to
absorbed water and conducting protons.

In our `united atom' approximation for Nafion, the  ether oxygens and
sulfur atoms are treated individually, while the  fluorocarbon groups
are consolidated as a single particle, as are the three oxygens of the
sulfonate \cite{article-1,article-2,vishnyakov2001}.  
The fluorocarbon groups, the sulfonate
oxygens, and the sulfur atoms  are modeled as single Lennard-Jones
(LJ) particles with a diameter $\sigma$= 0.35 nm. 
The protons carry the full formal
charge of $Q_{p}=+e$, the sulfur atoms have  $Q_S=+1.1e$, and the
combined triplet  of oxygen atoms carries $Q_{O_3}=-2.1e$. The partial
charges of the ether oxygens and the fluorocarbon LJ particles are 
set to zero.  Depending on whether the membrane  is dry or
hydrated,  two different representations have been used for the 
sulfonate head groups. For dry membranes, we implement 
an attached-proton model,  also called a dipole model for head
groups, which was extensively analyzed in our previous paper
\cite{article-1}. Though the attached-proton model does not
allow for proton diffusion,  it is considered as  a good starting
point for a step-by-step exploration of nanophase morphology in
PEM materials. For the hydrated ionomer we assume a detached-proton
model \cite{article-2,allahyarov-stretch},  where the protons diffuse
freely in the system, where they interact  with ionized head 
groups and water molecules.

The configurational part of the coarse-grained Hamiltonian for the
attached and detached proton models is
a combination of Coulomb interactions, non-bonded, and bonded
interactions between all the 
ionomer components: 
\begin{equation}
U_{\rm total}=U_{LJ}+U_{Q}+U_{\rm bond}+U_{\rm angle}+U_{\rm dihedral} \,\, .
\label{eq:1}
\end{equation}
Here  $U_{\rm total}$, $U_{LJ}$, $U_{Q}$, $U_{\rm bond}$, $U_{\rm angle}$
and $U_{\rm dihedral}$ are the total potential energy and its
Lennard-Jones, electrostatic, bond-stretching (bond-length term),
angle bending (bond-angle term) and dihedral angle components,
respectively: 
\begin{equation}
U_{LJ}(r)=4\varepsilon_{LJ}\sum_{i>j} \left( (\sigma/r_{ij})^{12}- a
  (\sigma/r_{ij})^{6} \right) ,
 \label{eq:2}
  \end{equation}
\begin{equation}
U_{Q}=\sum_{i>j}\frac{Q_{i}Q_{j}}{\epsilon r{ij}},
 \label{eq:3}
  \end{equation}
\begin{equation}
 U_{\rm bond}(r)=\frac{1}{2} \sum_{\textrm{all bonds}} k_{b}(r-r_{0})^{2},
 \label{eq:4}
  \end{equation}
\begin{equation}
U_{\rm angle}(\theta)=\frac{1}{2}\sum_{\textrm{all angles}}
k_{\theta}(\theta-\theta_{0})^{2},
 \label{eq:5}
  \end{equation}
\begin{equation}
U_{\rm dihedral}(\alpha)=\frac{1}{2}\sum_{\textrm{all dihedrals}}
k_{\alpha}\left(1-d \cos(3\alpha)\right) .
 \label{eq:6}
\end{equation} 
In  Eq.(\ref{eq:2}) the LJ interaction
coefficient $\varepsilon_{LJ}$ in units of $k_{B}T$ was chosen to be
0.33.  The parameter $a$ in the LJ term is 1 for hydrophobic-hydrophobic
(HH) interactions, and 0.5 for hydrophobic-hydrophilic
(HP) interactions and  hydrophilic-hydrophilic (PP) interactions.
In the latter case only the repulsive part of the LJ potential (a shifted
LJ potential for $r<1.1224\sigma$) has been considered.
 In  Eq.(\ref{eq:3}),  $Q_i$ and $Q_j$ are the electrostatic charges of the
 two interacting elements, which can be sulfur atoms, oxygen triplets,
 protons, or  the hydrogens or oxygen of the water molecules, and
 $\epsilon$ is  the dielectric constant of the ionomer.
In Eqs.(\ref{eq:4})--(\ref{eq:6}) the following force-field parameters 
have been used: the equilibrium 
bending angle  $\theta_{0}=110^{0}$, the equilibrium bond length
$r_{0}=0.44\sigma$, the bending force constant
$k_{\theta}=120\frac{\textrm{kcal}}{\textrm{mol }\textrm{deg}^{2}}$ 
and the stretching force constant $k_{b}=7 \times
10^4\frac{\textrm{kcal}}{\textrm{mol}\textrm{(nm)}^{2}}$. 
The dihedral angle parameters were $d=-1$ $(+1)$ and $k_{\alpha}=10.8 k_BT$
$(k_{\alpha}=3.7k_BT)$ for the backbone (sidechain) segments. 

 The dielectric properties of the coarse-grained material are represented by
   a distance-dependent dielectric function, 
\begin{equation}
\epsilon(r)=1 + \epsilon_{B}(1-r/\sigma))^{10}/(1+(r/\sigma)^{10}),
\label{epsilon}
\end{equation}
 where $\epsilon_{B}$ is  the bulk dielectric constants of the
 ionomer.  Usually a uniform  permittivity $\epsilon=1$,  or
 equivalently  $\epsilon_{B}=0$ in Eq.(\ref{epsilon}),  is  accepted
 in  {\it ab initio}  quantum-mechanical simulations, 
where all the ionomer atoms are explicitly taken into account. 
 Since the coarse-grained approach neglects the atomistic structure
 of the ionomer monomers,  additional approximations for the dielectric
permittivity have to be made to account for the polarization effects of
 the ionomer monomers as a response to the strong electrostatic fields
 of the sulfonate groups and protons. 
To be accurate, $\epsilon(r)$ should depend upon the atom types
and the absolute values of all the explicit coordinates. The problem,
of course, is that the specific form of $\epsilon(r)$ is not known.
For the bulk dielectric constant $\epsilon_{B}$ we use
$\epsilon_{B}$=8, which is appropriate  to the dielectric permittivity of Nafion
as measured in high-frequency studies \cite{paddison-1998-epsilon} and
differential scanning calorimetry \cite{lu-2008-epsilon}, 
  and from first-principle calculations \cite{tsampas-2007}. 

Taking into account the fact that the sulfonic acid tips of sidechains are
hydrophilic, and the remaining part of sidechains, as well as the
backbone polymer, are hydrophobic
\cite{Paddison1,paddison-review-1},  we use the following notation to describe
the polymer architecture:  $n_{1}H+n_{2}P$ for sidechains and
$n_{3}H$ for the backbone segments. Here $n_{1}$ is the number of
hydrophobic monomers per sidechain,  $n_{2}$ is the number of
hydrophilic monomers per sidechain, and  $n_3$ is the number
of backbone monomers between two adjacent sidechains. 
 The total number of sidechain monomers per
pendant chain is $n_1+n_2$ in the detached proton model, and
$n_1+n_2+1$ in the attached-proton model. 
The key variables that describe our model ionomer system are listed in
 Table~\ref{table-1}.

\section{Simulation details}
\label{sectionsimul}
Extensive coarse-grained molecular dynamics simulations
 were performed to investigate the swelling properties of
 sulfonate multiplets at four different solvation parameters $\lambda$
 and two distinct  sulfonate molar concentrations $\eta$. 
The parameter  $\eta$  is defined
as  $\eta \equiv \left( N_S/N_0 \right) V$, where $N_S$ is
 the number of sidechains in the volume $V=L^3$ of 
 the simulation cell, and $N_0$ is Avogadro's  number. 
Whereas in experimental studies the parameters $\eta$ and $\lambda$
are  coupled to each other \cite{gebel-moore-2000}, in
 numerical simulations both quantities can be changed independently.  
A series of simulation runs are summarized in Table \ref{table-2}. 
The molar concentrations $\eta_1$=0.8 mol/l
and $\eta_2$=1.5 mol/l correspond to the  backbone segment
lengths $n_3$=50 and $n_3$=20 respectively. Varying the parameter
 $\eta$, i.e. varying the chain volume per SO$_{3}^{-}$ group, is in
 some ways equivalent to simulating materials with different
 equivalent weights  \cite{gebel-moore-2000}. For convenience, we will
 refer to the membrane with sulfonic 
 molar concentration $\eta$ as `membrane $\eta$'. 
 In most simulations of Nafion-like ionomers
 the parameter $n_3$ is usually 
 varied between 14 and 18. Shorter segments with
$n_3$=10 have been considered in the atomistic simulations of
 Ref.~\cite{paddison2}, and longer segments $n_3$=30 in the
 coarse-grained approaches of  Ref.~\cite{vishnyakovDPD}. 
In the latter case the  nearest-neighbor distance
 between the sulfonate multiplets is large. Thus our choice of a 
 larger $n_3$ makes  possible the investigation of the solvation properties of
 single multiplets  in slightly hydrated membranes. 

There were $N=(n_{1}+n_{2}+n_{3})\times N_{S}$ 
polymer monomers in the simulation box of length $L$. All simulations
were carried out for both the $n_{1}=7$ and $n_{2}=2$ sidechain
architectures. The number of sidechains $N_S$ was 500 for membrane
$\eta_1$ 
and 1000 for membrane $\eta_2$. The negative charges of $N_S$
sulfonate groups   were compensated by $N_S$ positive protons
to  guarantee an overall charge neutrality in the system. Simulations
with explicit water include an additional  $3\times\lambda \times N_S$ 
water charges, as we are using the SPC solvent model
\cite{water_parameters,jang2004}.  
The box size $L$ was systematically increased from  $L=30\sigma$ to
$L=32.5\sigma$ when the water content $\lambda$ was increased from
$\lambda$=0 to $\lambda$=5 in order to keep the density of the hydrated
membrane constant.

One of the main challenges in generic ionomer simulations is the
fact that the ionomer molecule is quite stiff at ambient temperatures and low
humidity conditions. In experimental studies a fast  ionomer
equilibration is usually achieved through different pretreatments
protocols, such as a soaking in a solvent or  high-temperature annealing. 
These steps  improve the sidechain kinetics and decrease the
barrier between the trapped metastable states and the low-lying states
at the global minimum in free energy.  
Overall, a full equilibration, even after these pretreatment steps,
takes hours or days, a time span that is far beyond the feasible 
simulation times of several nanoseconds in typical molecular dynamics
runs. 
To overcome this obstacle  we  implemented the following
artificial steps \cite{artificial-step}:\\
a) the sidechains were temporarily detached from the backbone
skeleton, a technique that has been successfully applied in
Refs.~\cite{article-2,allahyarov-stretch,rivin-vishnyakov,rivin2004,elliott1999}, \\
b) the skeleton was cut into smaller segments of length $n_3$.\\
The resulting `fragmented' ionomer reaches the  
 equilibrium state very fast because of the increased diffusive
 movement of its segments. Typical MD runs of 500 ps duration in the
 $NVT$ ensemble were enough to fully equilibrate the simulated
 system. The system temperature $T$ was controlled 
by coupling the ionomer to a Langevin thermostat with a friction coefficient
$\gamma=10$ps$^{-1}$  and a Gaussian white-noise force of strength $6k_{B}T\gamma$.
The equations of motion were integrated using the velocity Verlet
algorithm with a time step of 0.25 fs. We also imposed standard periodic
boundary conditions to our system, thus filling space with translational
replications of a fundamental cell. Long-range electrostatic interactions
were treated using the Lekner summation algorithm \cite{Lekner}.

 In the next stage of the simulations, the ionomer segments were reassembled
 back into a branched chain characterizing the original Nafion-like ionomer. 
This was achieved by a  simultaneous  introduction of  bonds and
 angular constraints 
 between the ends of backbone segments unifying them into a single and
 long backbone chain.  Similar bond and angular
 constraints were introduced between the fluorocarbon tail monomers
 of detached sidechains and the  median section 
monomers of backbone segments. To avoid the formation of unphysical
 star-like branches only a single occupancy of backbone attachment
 sites was permitted. The simulations were then resumed for another
 few hundred picoseconds until  a new equilibrium state was reached.
 Then the  statistically averaged  quantities of interest were
 gathered during the next 3ns--5ns of the long production runs.

\section{Simulation results}
\label{section-results}
\subsection{Multiplet formations in dry and solvated membranes}
\begin{figure}
\includegraphics*[width=0.9\textwidth]{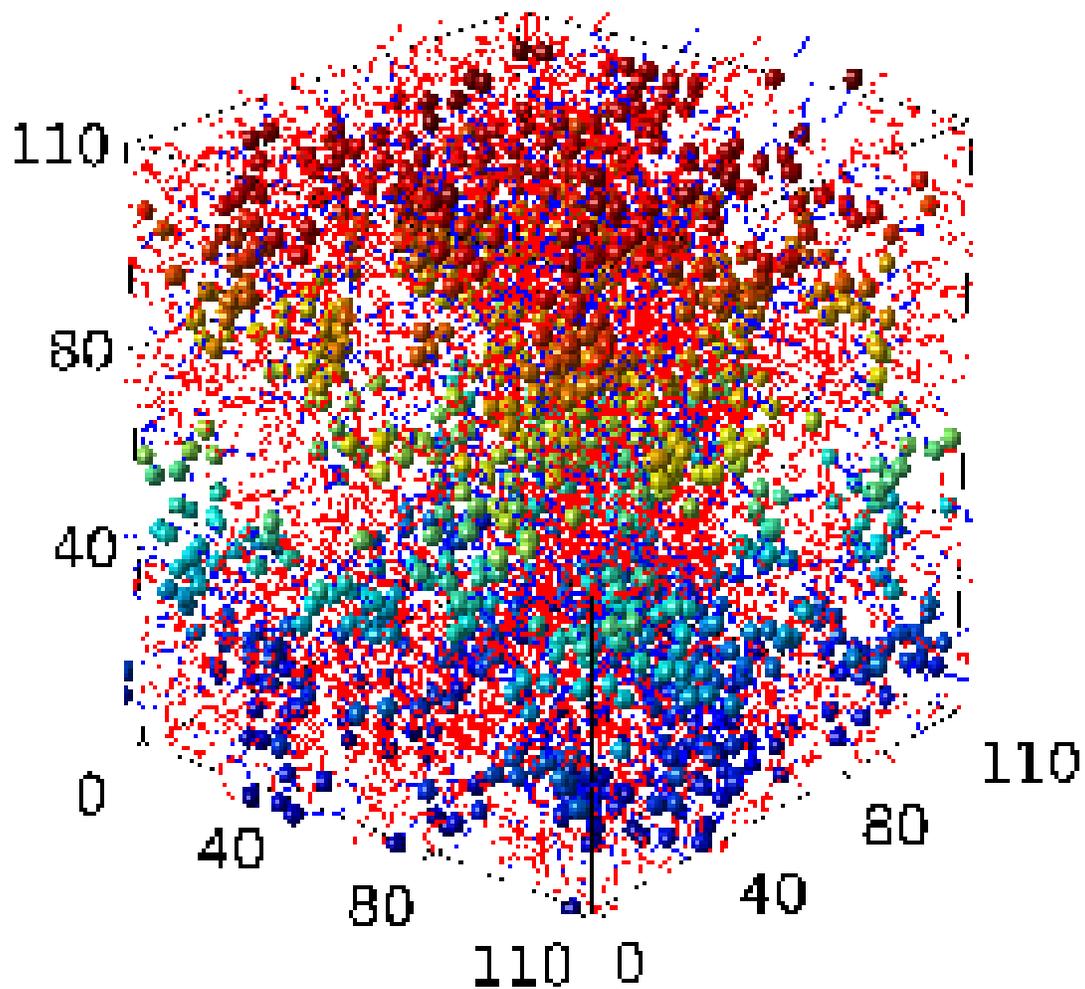}
\caption{(Color online) A typical snapshot of  hydrated membrane
 $\eta_2$ from Run~3. The spheres represent the  end-group oxygens of the
 sidechains. The  polymer is shown by red lines.  Different bead colors
 correspond to different bead altitudes, with a blue color for
 low-altitude beads (at the bottom of simulation box) and a red color
 for high-altitude beads (at the top of simulation box). The size of
 all structural elements is schematic rather than space filling. The
 water molecules and protons are not shown for the sake of clarity.
\label{fig2-snapshot}} 
\end{figure}

\begin{figure}
\includegraphics*[width=1.\textwidth]{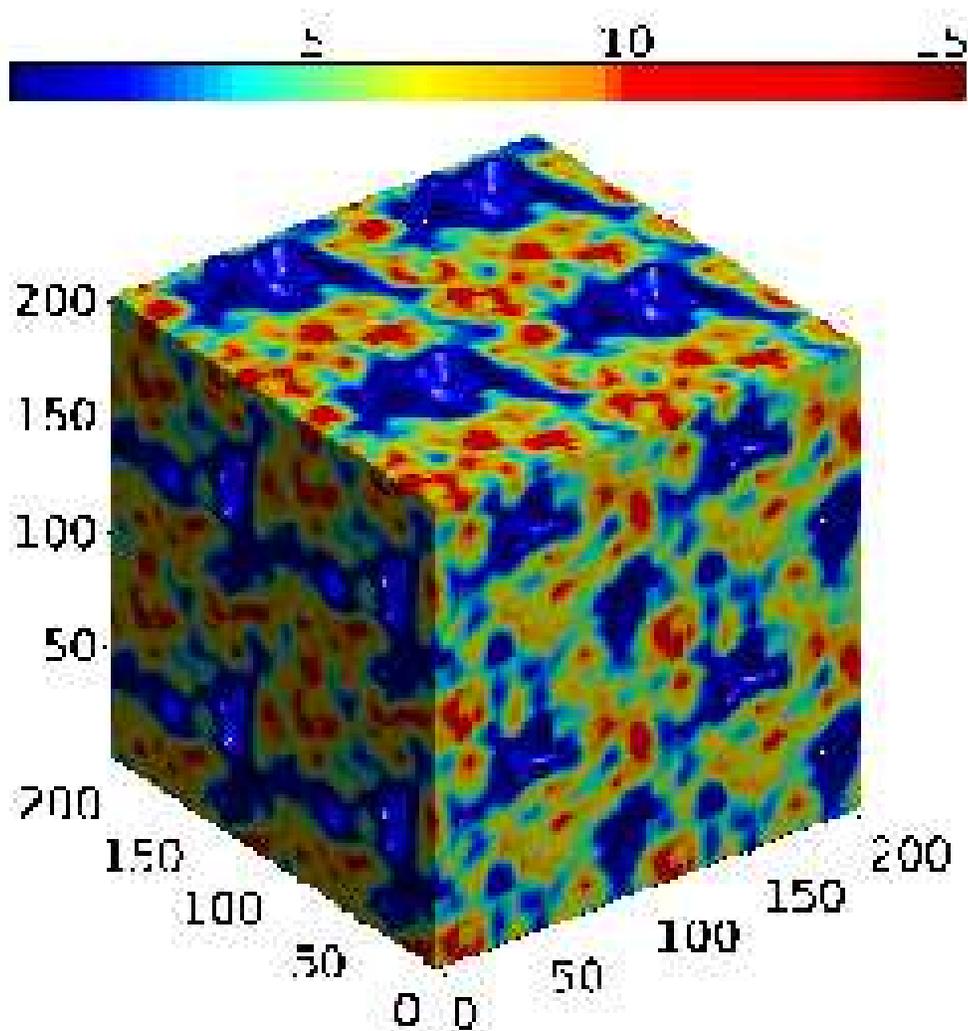}
\caption{(Color online) 3D density $\rho(\vec r)$ of the hydrophobic part of the
membrane  $\eta_2$ for Run~3. The color gradient from
  dark blue (black in printed version) to dark red (gray in printed
  version), corresponds to the variation of 
  membrane density from low to high values. The axis dimension
  is in \AA.  
\label{fig10-density-membrane}}
\end{figure}

\begin{figure}
\includegraphics*[width=1.0\textwidth]{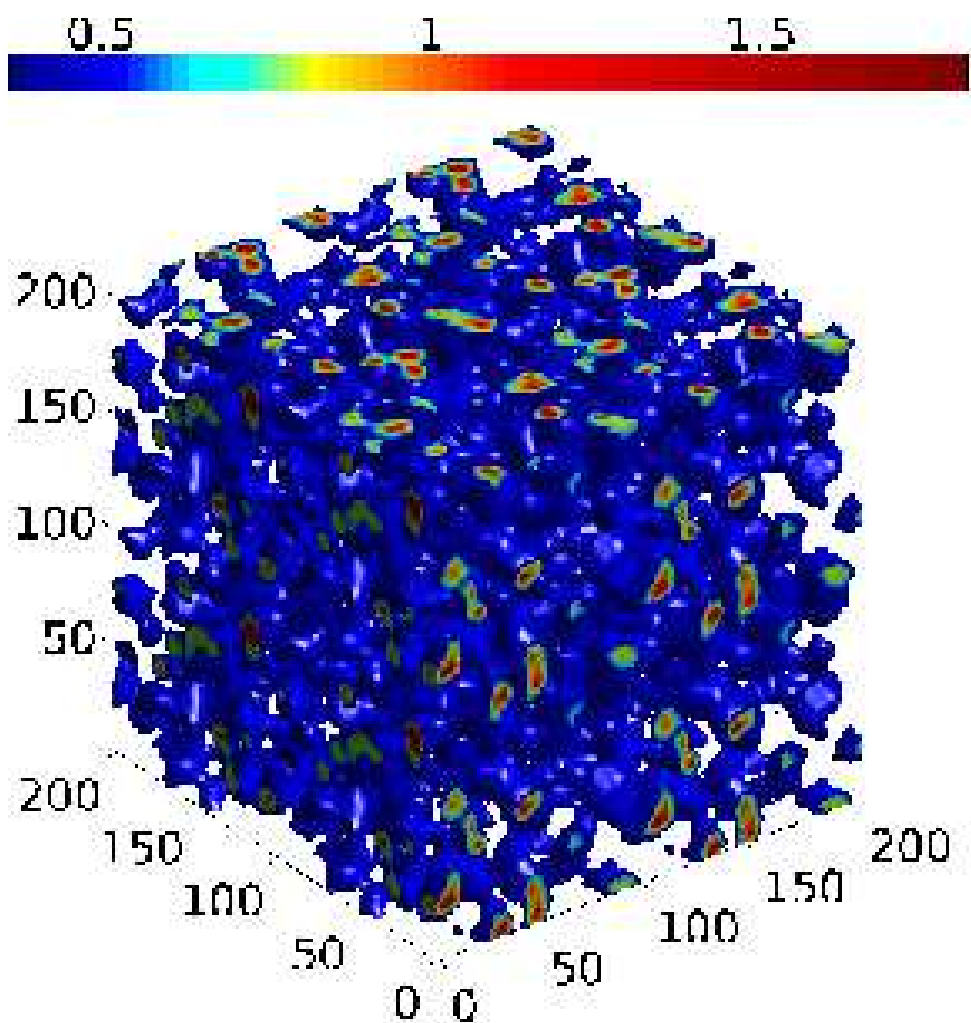}
\caption{(Color online) 3D density of 
  water channels for the membrane
  $\eta_2$ and  Run~3. The color gradient from
  dark blue (black in printed version) to dark red (gray in printed
  version) corresponds to the variation of  water
  density from low values  to high values. The axis dimension 
  is in \AA.
 \label{fig11-density-water}}
\end{figure}

A typical snapshot of a hydrated membrane from Run~3  
is shown in Figure~\ref{fig2-snapshot}. The backbone skeleton, plotted
as lines, creates a hydrophobic network with chaotically scattered
pores. These pores incorporate micelle-like clusters of sidechain
sulfonates (shown as spheres), which are filled with
water molecules and protons.  The number  densities $\rho(\vec r)$ of
the hydrophobic part of the ionomer and of the absorbed water,
averaged over  a 100 fs run, are shown  in 
Figure~\ref{fig10-density-membrane} and
Figure~\ref{fig11-density-water}, respectively. The density $\rho(\vec
r)$ corresponds to the probability of finding a particular membrane
component, hydrophobic monomer or hydrophilic water, at the point
$\vec r$ during a short simulation run.    
The quasi-regular 
network of  polymer skeleton with interconnected 
 hydrophilic pores changes its form slowly with time.  

\begin{figure}
\includegraphics*[width=0.6\textwidth]{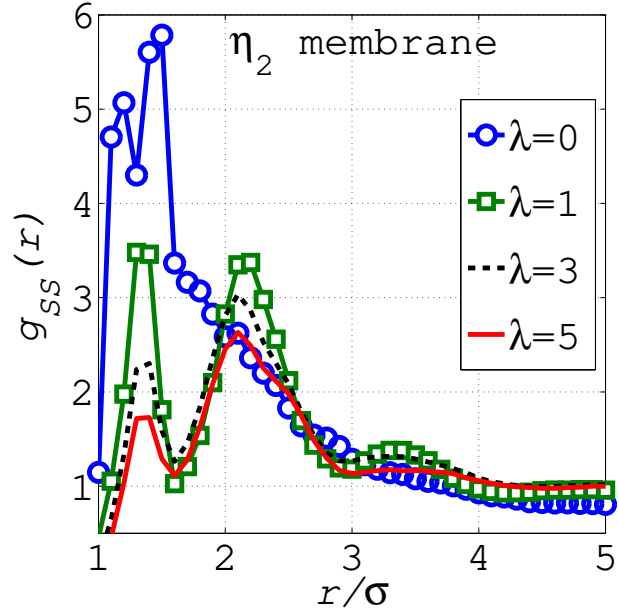}
\includegraphics*[width=0.6\textwidth]{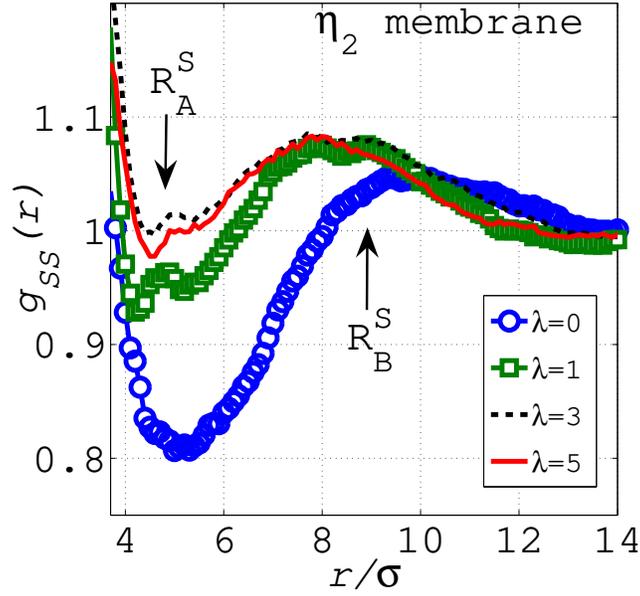}
\caption{(Color online)
Sulfonate-sulfonate pair correlation function $g_{SS}(r)$ for
membrane $\eta_2$ as a function of sulfur-sulfur separation
distance $r$ for Runs~1--4 from
Table~\ref{table-2}. Solid line with circles - Run~1, solid
lines with squares- Run~2, dashed line- Run~3, full line- Run~4. The
bottom figure shows in detail the long-range tail of $g_{SS}(r)$ used to
determine the average multiplet size $R_A^S$ and the  separation
distance between the sulfonate multiplets $R_B^S$. The calculated values
for  the parameters $R_A^S$ and $R_B^S$ are given in Table~\ref{table-eta}.
  \label{fig3-gss-a}}
\end{figure}

The structure of the sulfonate multiplets was probed through
the calculation of the sulfonate-sulfonate pair correlation function,
\begin{equation}
g_{SS}(r)=\frac{V}{N_{S}}\frac{dn_{S}(r)}{4\pi r^{2}dr} \,\,\,\,\,   .
\end{equation}
Here $dn_{S}(r)$ is the number of sulfurs located at the distance
$r$ in a shell of a thickness $dr$ from a fixed sulfur atom.
The function $g_{SS}(r)$ indicates the probability of finding two sulfonate
monomers at a separation distance $r$ averaged over the equilibrium
trajectory of the simulated system.
Simulation results for $g_{SS}(r)$ for Runs~1--4 from Table~\ref{table-2}
 are shown  in Figure~\ref{fig3-gss-a}. The dry multiplets   
 have no detectable internal structure except the strong maximum  at
 $r\approx$1.4$\sigma$. In hydrated membranes  the
 correlation function $g_{SS}$  shows shell-like oscillations, a
 recognizable fingerprint of solvation shells. The first maximum of
 $g_{SS}(r)$ corresponds to the closest-approach configuration between
 neighboring sulfonates. The second peak of
 $g_{SS}(r)$  stems from  a configuration where two neighboring
 sulfonates are separated by single proton or water molecule. 
 Finally, the third peak of $g_{SS}(r)$ is related to configurations with
 more than one proton or water molecule  between sulfonates. 
The dependence of the intensity of the correlations between the head groups  on
water content $\lambda$ is due to the dielectric screening properties
of water: the more the water content in the membrane, the weaker the
sulfonate-sulfonate interactions. Similar results have been reported
in the simulation results of
Refs.~\cite{cui-2007,cui-2008,brunelo-2009,spohr-2004}. The greatly
reduced intensity of the first peak of $g_{SS}(r)$ at $\lambda=5$ can
be understood as the onset of the improbability of the closest-approach
sulfonate-sulfonate configurations in hydrated
multiplets. 

We determine the size of a multiplet as the position of the global
 minimum $R_A^S$ (also known as the radius of the first coordination sphere)  
 of the pair correlation functions $g_{SS}$ in
 Figure~\ref{fig3-gss-a}. This position depends on the membrane
 hydration level $\lambda$, and can be used to calculate the
 number of head groups  $\chi_S$ inside the 
multiplet according to the following relation 
\begin{equation}
\chi_S = \frac{N_S}{V} \int_0^{R_A^S} g_{SS}(r)  4 \pi r^2 dr .
\label{population}
\end{equation}
The calculated values for the parameters $R_A^S$ and $\chi_S$ are given in
Table~\ref{table-eta} for the membranes
$\eta_1$ and $\eta_2$. There is a clear
indication of the fact that the multiplets shrink in size at the onset
of membrane solvation, which corresponds to the transition from Run~1
to Run~2. This shrinking, which is not in accord with the
classical theories of cluster swelling in ionomers, 
is accompanied by  a multiplet splitting into smaller parts. For
instance, the dry multiplet in the membrane $\eta_2$ has a size
$R_A^S=5\sigma$ and consists of $\chi_S=22$ head groups. Following
hydration by a water content as low as $\lambda=1$, this multiplet
effectively splits into two smaller parts of size $R_A^S=4.2\sigma$,
each of them consisting of only 13 head groups. These small multiplets
will consequently swell, keeping the number of their sulfonate
population constant, when additional water is absorbed into the
membrane. The swelling radius is largely determined by the competition
between two different internal energies, the elastic energy of the 
backbone material and the electrostatic energy of the pendant
groups. 

\begin{figure}
\includegraphics*[width=1.\textwidth]{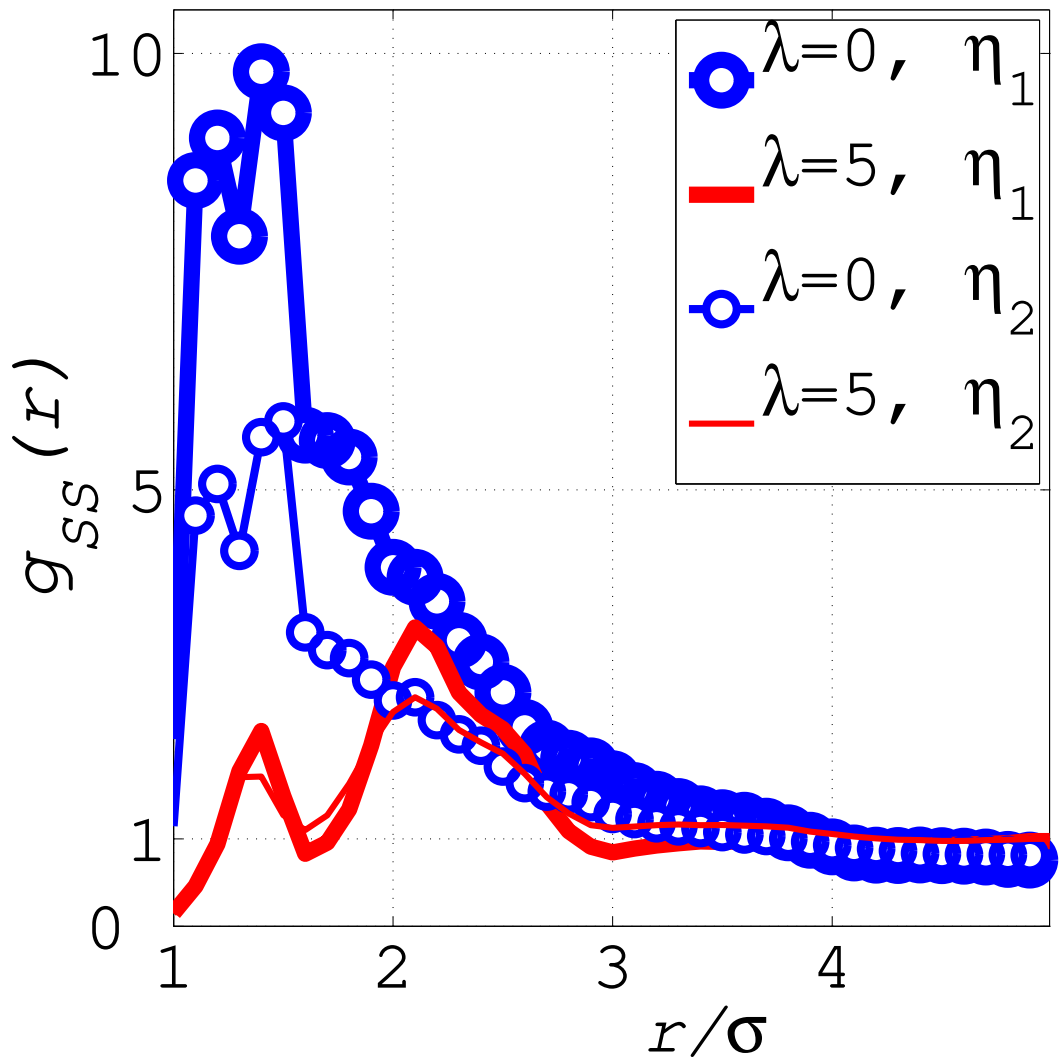}
\caption{(Color online)
Sulfonate-sulfonate pair correlation function $g_{SS}(r)$ for
the membranes $\eta_1$ and $\eta_2$ as a function of the sulfur-sulfur separation
distance $r$. Lines with symbols- Run~1, full lines- Run~4. 
  \label{fig3-gss-b}}
\end{figure}

The pair correlation functions  $g_{SS}(r)$ for two different sulfonate
concentrations $\eta_1$ and $\eta_2$ are plotted 
in  Figure~\ref{fig3-gss-b}. When the parameter $\eta$ decreases,
(as seen by a comparison of the thin and thick lines in Figure~\ref{fig3-gss-b}),
  the intensity of sulfur-sulfur correlations
 increases.  This effect stems from the interplay between the
 electrostatic screening length $l_D$ and the average separation
 distance $\overline l$ between the sulfonates. For an 
 ionomer with a high sulfonate concentration $\eta$, one generally has
 $l_D<\overline l$, and  thus the electrostatic 
 correlations between the sulfonates are negligible. In this case a
 nanophase separation in the membrane is possible only due to the
hydrophobic/hydrophilic immiscibility between the  backbone and 
sidechain segments of the membrane.  In the opposite case, when $\eta$ is
small and $l_D>\overline l$, the Coulomb correlations become 
sufficiently strong to force the sulfonates to form compact multiplets.

\subsection{Separation distance between multiplets}
\begin{figure}
\includegraphics*[width=1.\textwidth]{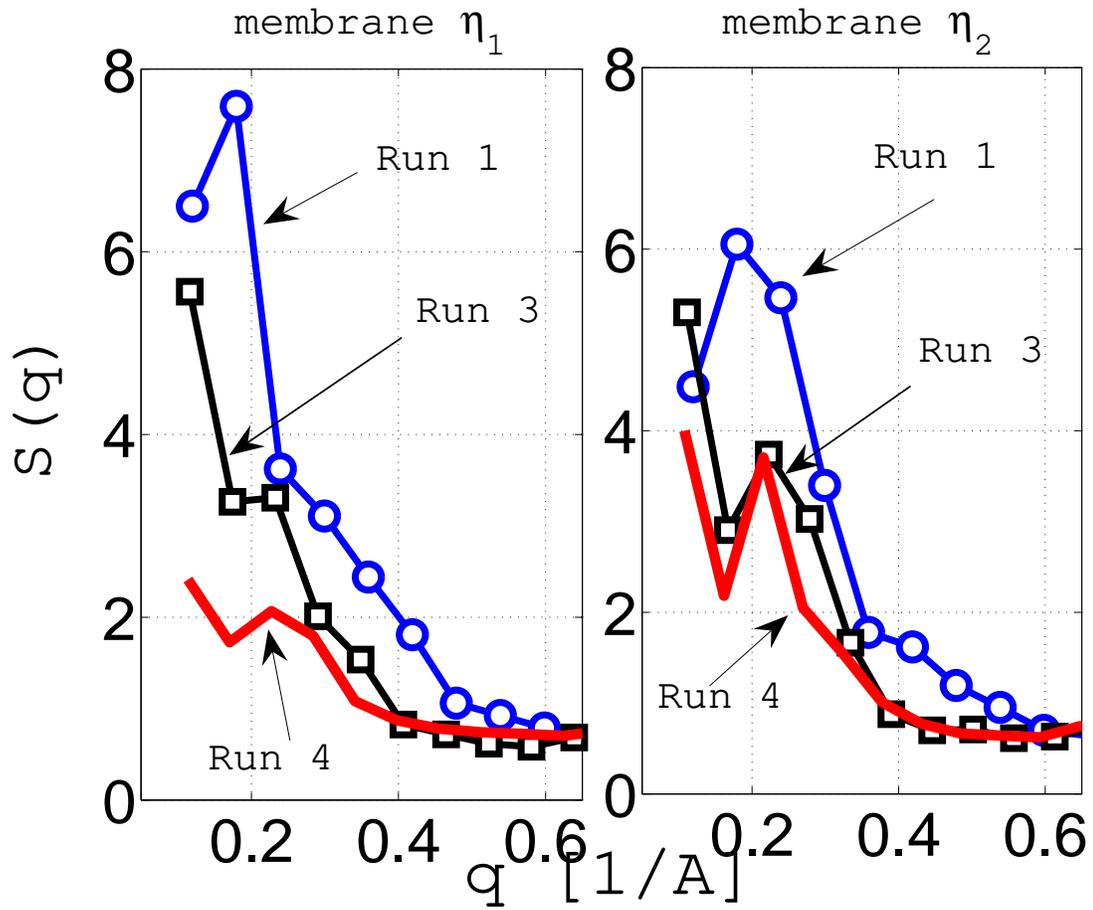}
\caption{(Color online) Small angle ionomer peak region of
  sulfonate-sulfonate structure factor $S(q)$ 
for the membranes $\eta_1$ and $\eta_2$. Line with circles-
  Run~1, line with squares- Run~3, full line-  Run~4.
  The step size $\delta q=2\pi/L \approx 0.05$\AA$^{-1}$ defines
the resolution along the {\it x}-axis.  
 \label{fig4-structure}}
\end{figure}

It is a well established fact that the nearest-neighbor separation
distance between the multiplets and the 
connectivity of multiplets into a network of hydrophilic pathways 
are the main contributing factors to the transport properties of ionomers.
 The typical multiplet-multiplet nearest-neighbor distances can
 be directly deduced from the density-density correlations in the
network of  head groups by consideration of the structure factor, 
\begin{equation}
S(\vec{q})=N_{S}^{-1}\left\langle
\left[\sum_{i=1}^{N_{S}}\cos\left(\vec{q}\vec{r_{i}}\right)\right]^{2}
+\left[\sum_{i=1}^{N_{S}}\sin\left(\vec{q}\vec{r_{i}}\right)\right]^{2}\right\rangle . 
\label{sk}
\end{equation}
When there is no preferential ordering of the hydrophilic domains in the
membrane, the structure factor of the sulfonates  is isotropic, and hence depends only
on the modulus $q=\vert{\vec q} \vert $ of the wave-vector. 
The calculated structure factors $S(q)$ for the dry and hydrated
 membranes are presented in Figure~\ref{fig4-structure}.  The
 ionomer-peak position  in the low $q$-region corresponds to the
 length of the  density-density  correlations  $\overline R=2\pi/q$ of
 sulfonates  \cite{Gebel,gebel-moore-2000,elliott2006,young2002}.  

 The nearest-neighbor distance between the multiplets can also be
 deduced, though less precisely, from
 the position of the long-range maximum $R_B^S$ of
 the pair correlation functions $g_{SS}(r)$ in Figure~\ref{fig3-gss-a}. 
 As in the multiplet splitting effect, corresponding to the 
reduction in the  multiplet size $R_A^S$ at the onset of membrane
solvation, the nearest-neighbor distance  $R_B^S$
 also decreases to smaller values according to the results of
Run~1 and Run~2. This is a consequence of the increase in the
 multiplet population $\xi=N_S/\chi_S$. For example, the
 number of multiplets in the  membrane $\eta_1$ increases from $\xi$=30
 in the dry membrane to $\xi$=50 in the hydrated membrane.  
The calculated values for $\overline R$ and $R_B^S$, seen in  
Table~\ref{table-eta}, match each other perfectly.
 We note that the increase of the average multiplet separation distance
 $\overline R$ for Runs~2--4 is a clear sign of membrane
 swelling, which  is in accord with the  results of
 Refs.~\cite{sacca-2006,cui-2007,gebel-moore-2000}.

\subsection{Swollen  multiplets inside a water shell}
\begin{figure}
\includegraphics*[width=1\textwidth]{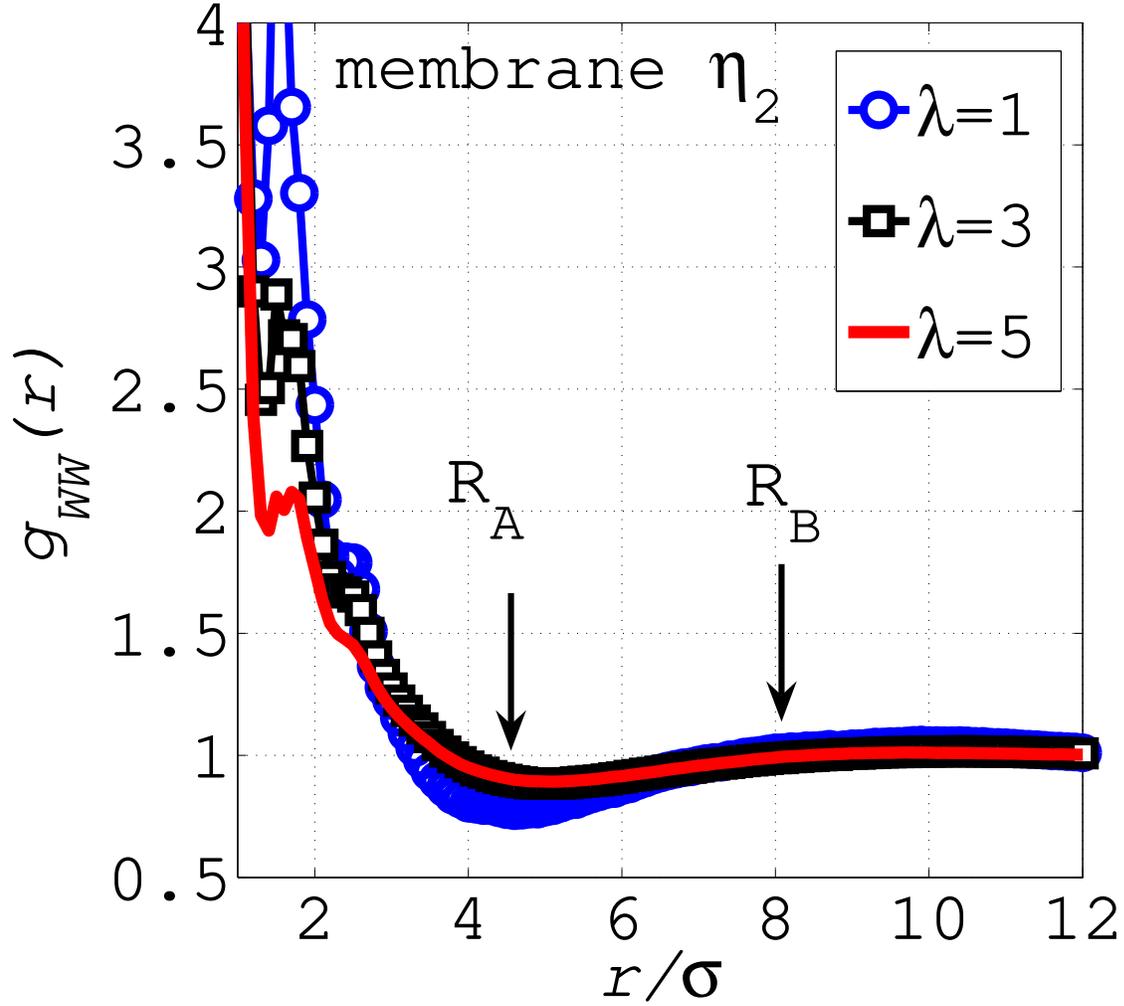}
\caption{(Color online)
Water-water pair correlation function $g_{WW}(r)$ for
the membrane $\eta_2$ as a function of the water-water separation
distance $r$ for Runs~2--4 from Table~\ref{table-2}. Solid line with
circles - Run~2, solid 
lines with squares- Run~3,  full line- Run~4. The
arrows  show the average multiplets size $R_A^W$ and the  separation
distance  $R_B^W$ between the water clusters. The calculated values for
the parameters $R_A^W$ and $R_B^W$ are given in Table~\ref{table-eta}.
  \label{fig3-gww}}
\end{figure}

The pair correlation and the structure factor analysis,
implemented in the previous subsection, can be also exploited to
examine the water clustering features in hydrated membranes for
Runs~2--4.  We calculate the size of the water cluster $R_A^W$ from
the water-water correlation $g_{WW}(r)$ shown in
Figure~\ref{fig3-gww}. The nearest-neighbor cluster separation
distances $R_B^W$ were evaluated  from the water-water structure
factors. The calculated values for both parameters are given in 
Table~\ref{table-eta}. There is good 
agreement between the water-water and the sulfonate-sulfonate multiplet 
nearest-neighbor  distances $R_B^W$ and $R_B^S$.  This is an indirect
verification of the fact that the ionomer cluster is a mixture of
sulfonates and absorbed water molecules. The distribution  of water
molecules inside the ionomer cluster can be analyzed by comparing the
water cluster size $R_A^W$ with the sulfonate multiplet size $R_A^S$. 
Whereas  for the membrane $\eta_1$ there is an
excellent match between these two parameters, for the membrane $\eta_2$ 
the water clusters are consistently bigger than the sulfonate multiplets.    
Based on this result we conclude that a part of the total water
 loading per multiplet in fact exists outside the multiplet
 boundaries. This `outer' water shell encapsulates the multiplet
 and facilitates the formation of  narrow water channels between the
 swollen multiplets. These channels, clearly seen in 
 Figure~\ref{fig11-density-water}, are the 
 pathways through which the ionomer absorbs more solvent upon
its  hydration. The water channels are also 
 an attractive place for the unclustered head groups, and assist
 the proton diffusion between neighboring multiplets. 
 We remark that  free bulk-like water would form in the interior of
 the multiplet  only at sufficiently high solvation levels $\lambda$
 \cite{choi2005}, a case not considered in this work.

The average number of water molecules $\chi_W$
per water cluster was calculated by using Eq.(\ref{population}) for
$g_{WW}(r)$. This parameter, together with the parameter describing
the water-per-sulfonate ratio $\chi_W/\chi_S$  
are given in Table~\ref{table-eta}. For the membrane $\eta_1$  we
obtain $\chi_W/ \chi_S=\lambda$, a predicted result for the ionomer
cluster with $R_A^W=R_A^S$. However for the membrane
$\eta_2$ the ratio $\xi_W/\xi_S > \lambda$. This unexpected result can 
be interpreted in the following manner: when the sulfonate
concentration approaches the percolation threshold for head-groups,
a fraction of the sulfonates are randomly
distributed between the existing multiplets. These bridging  sulfonates
cannot retain their full solvation shell with 
 $\lambda$ water molecules in the hostile
environment of hydrophobic backbones.  The excess water
molecules stripped from these `bulk' sulfonates are consequently
redistributed  between the existing multiplets. This leads to the
formation of an outer solvent shell around each multiplet.

\subsection{Swelling-induced ionomer deformation}
The polymer backbone and sidechains sustain  conformational changes
when the membrane swells. Two different types of deformation, an
elongation (stretching) deformation and a coiling (frustration)
deformation of polymer chains can be  conveniently resolved  using the
probability distribution $P(\alpha)$ of  the dihedral angle along the
polymer chains.   

\begin{figure}
\includegraphics*[width=1.\textwidth]{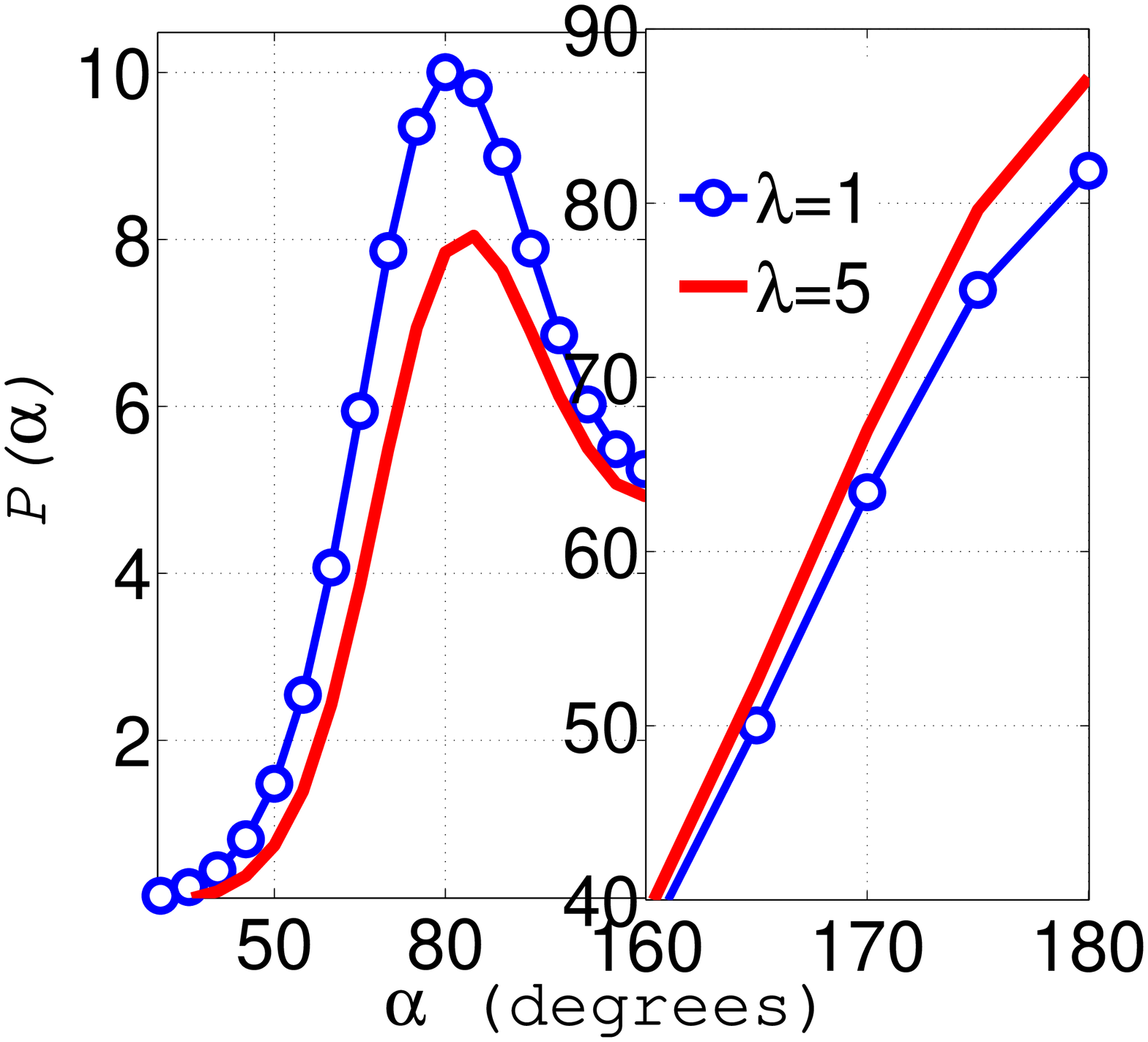}
\caption{(Color online) 
The probability distribution $P(\alpha)$ of the dihedral angle
  along the sidechain of membrane $\eta_1$,  and for Run~2 and Run~4. 
  The areas of the {\it gauche} conformation ($\alpha=82$ degrees)
  and the {\it trans} conformation ($\alpha=180$ degrees) are shown separately.       
 \label{fig5-dihedral}}
\end{figure}

\begin{figure}
\includegraphics*[width=1.\textwidth]{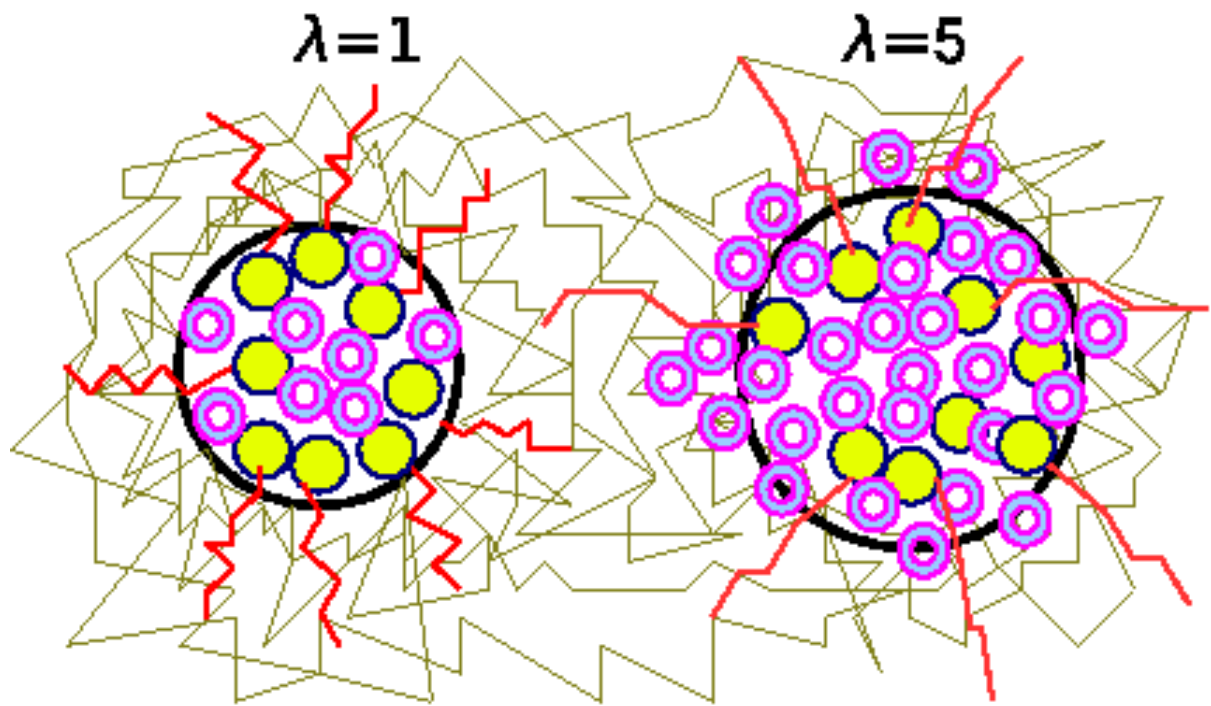}
\caption{(Color online) Schematic pictures explaining the sidechain  
  stretching-like relaxation as a result of multiplet swelling from $\lambda=1$ to $\lambda=5$.
The small hollow spheres are the water molecules, gray (yellow in
  online version) small spheres with attached tails are for
  sidechains, big spheres represent the  multiplets. 
 \label{fig7-a}}
\end{figure}

\begin{figure}
\includegraphics*[width=1.\textwidth]{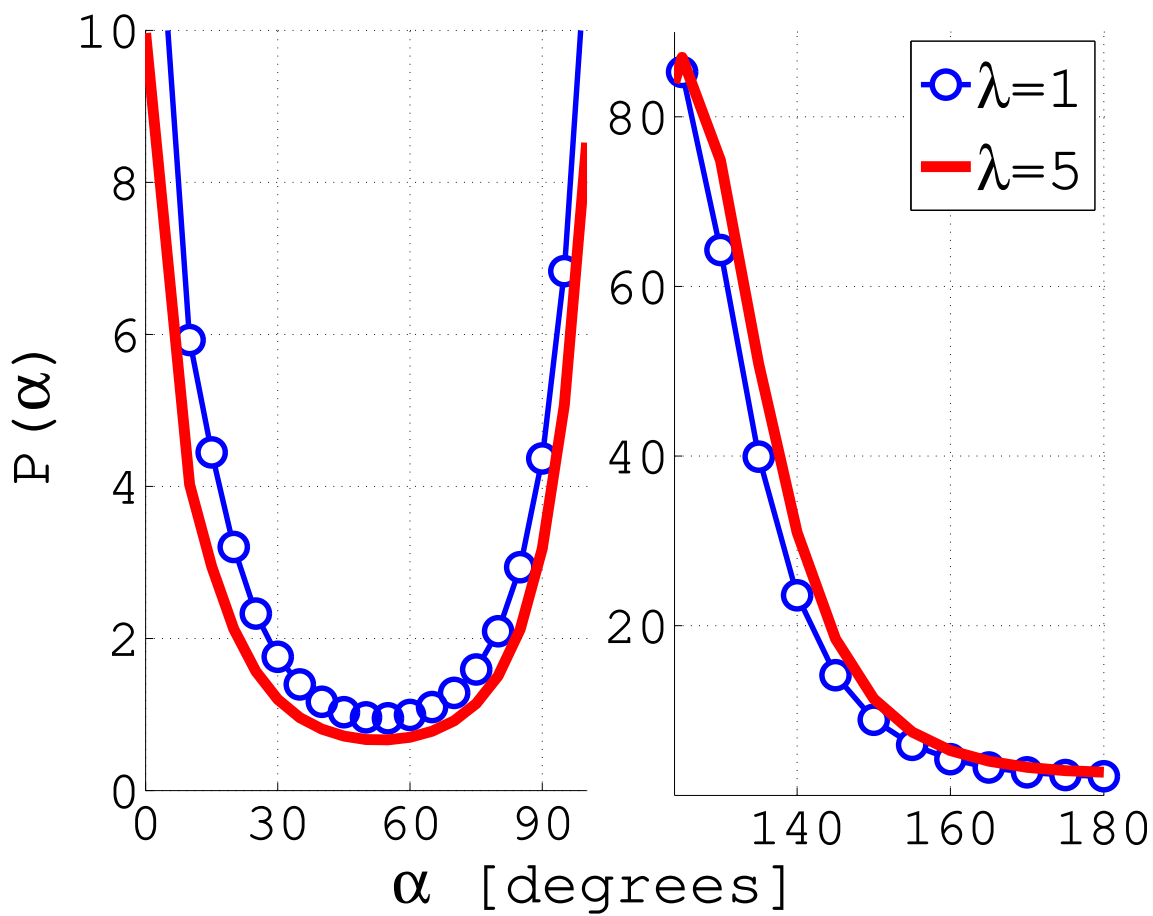}
\caption{(Color online) 
The probability distribution $P(\alpha)$ of the dihedral angle along the backbone of
 membrane $\eta_1$ for Run~1 and Run~4. 
  The areas of a  {\it cis} conformation ($\alpha=0$ degrees) and
 the {\it gauche} conformation ($\alpha=125$ degrees)  are shown separately.        
 \label{fig6-dihedral}}
\end{figure}

  The probability distribution $P(\alpha)$ of the dihedral angle along
 the sidechain is shown  in Figure~\ref{fig5-dihedral} for the membrane
 $\eta_1$. The sidechains have two  {\it gauche} ($\pm$82 degrees) and
 one {\it trans} conformations. When  the membrane absorbs water, the  
sidechains undergo a deformation in which a part of the  {\it gauche}
conformations  transform into {\it trans}
conformations. The overall effect of this structural deformation is a
 structural relaxation of the sidechains, perceived as a stretching --
 an impact schematically illustrated  in Figure~\ref{fig7-a}. 
We have also detected a similar stretching-like structural relaxation
 for the backbone segments. As seen from Figure~\ref{fig6-dihedral},
 the probabilities of the two {\it gauche} ($\pm$125 degrees) and
 single {\it cis} ($\pm$0 degrees) backbone conformations in the
 solvated membrane diminish when the hydration parameter $\lambda$
 decreases.

\begin{figure}
\includegraphics*[width=1.\textwidth]{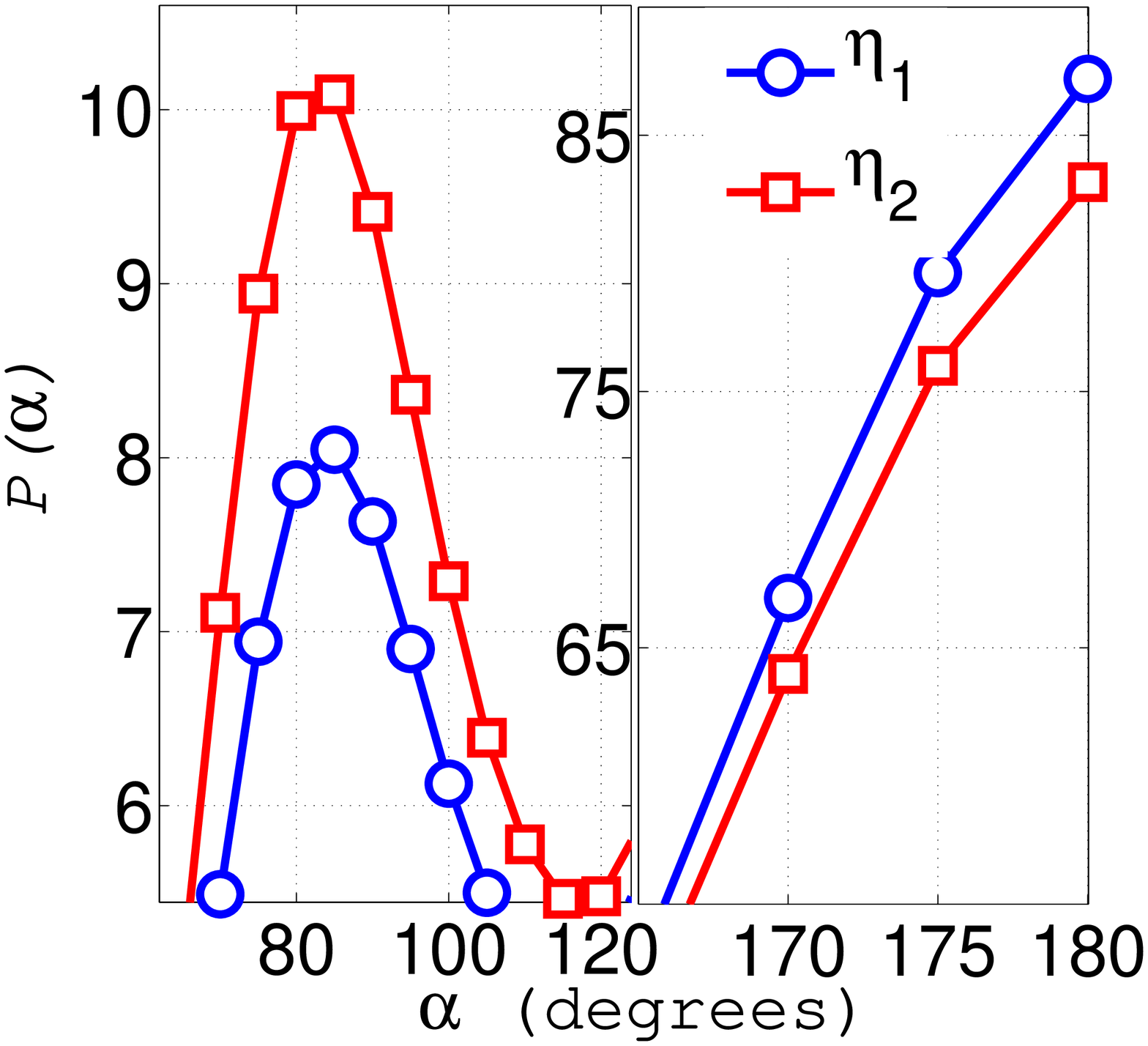}
\caption{(Color online)  The probability distribution $P(\alpha)$ of the dihedral angle
 along the sidechains of membranes $\eta_1$
  and $\eta_2$, and  for Run~4. 
 The areas of the {\it gauche} conformation ($\alpha=82$ degrees) and the {\it
  trans} conformation ($\alpha=180$ degrees) are shown separately.   
 \label{fig8-dihedral}}
\end{figure}

\begin{figure}
\includegraphics*[width=1.\textwidth]{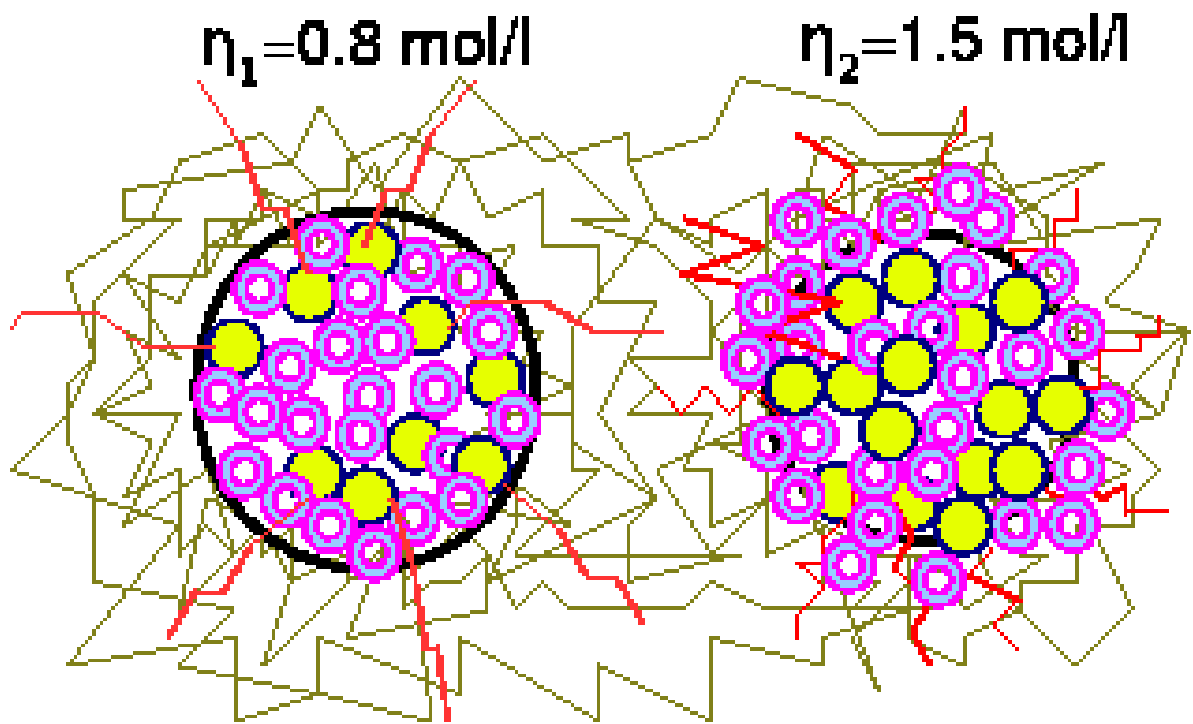}
\caption{(Color online) Schematic pictures explaining the differences
  between the stretching-like relaxation of sidechains for the membranes $\eta_1$ and
  $\eta_2$. The small hollow spheres are the water molecules, gray (yellow in
  online version) small spheres with attached tails are for
  sidechains, big spheres represent the  multiplets. 
 \label{fig7-b}}
\end{figure}

The extent of sidechain relaxation sensitively depends on 
 the sulfonate concentration $\eta$. In
Figure~\ref{fig8-dihedral} we compare
the  $P(\alpha)$ curves for the two membranes.  It is noticeable that
the sidechains are in a more relaxed state in the membrane $\eta_1$
 compared to the membrane $\eta_2$. This is a direct consequence of
 the fact that the smaller number of head groups inside the multiplet  provide
  a more relaxed configuration for sidechains compared to the case
 when a larger number of sulfonates are 
 immersed into a  smaller multiplet.  The dihedral frustration of
 sidechains in the high-sulfonate  membrane is schematically illustrated
 in   Figure~\ref{fig7-b}.  A similar dihedral frustration has also  been
 detected for the  backbone polymer: in the high-sulfonate membrane
 the  backbone segments adopt a more curly
 conformation.

\subsection{Proton Diffusion}
\begin{figure}   
\includegraphics*[width=1.\textwidth]{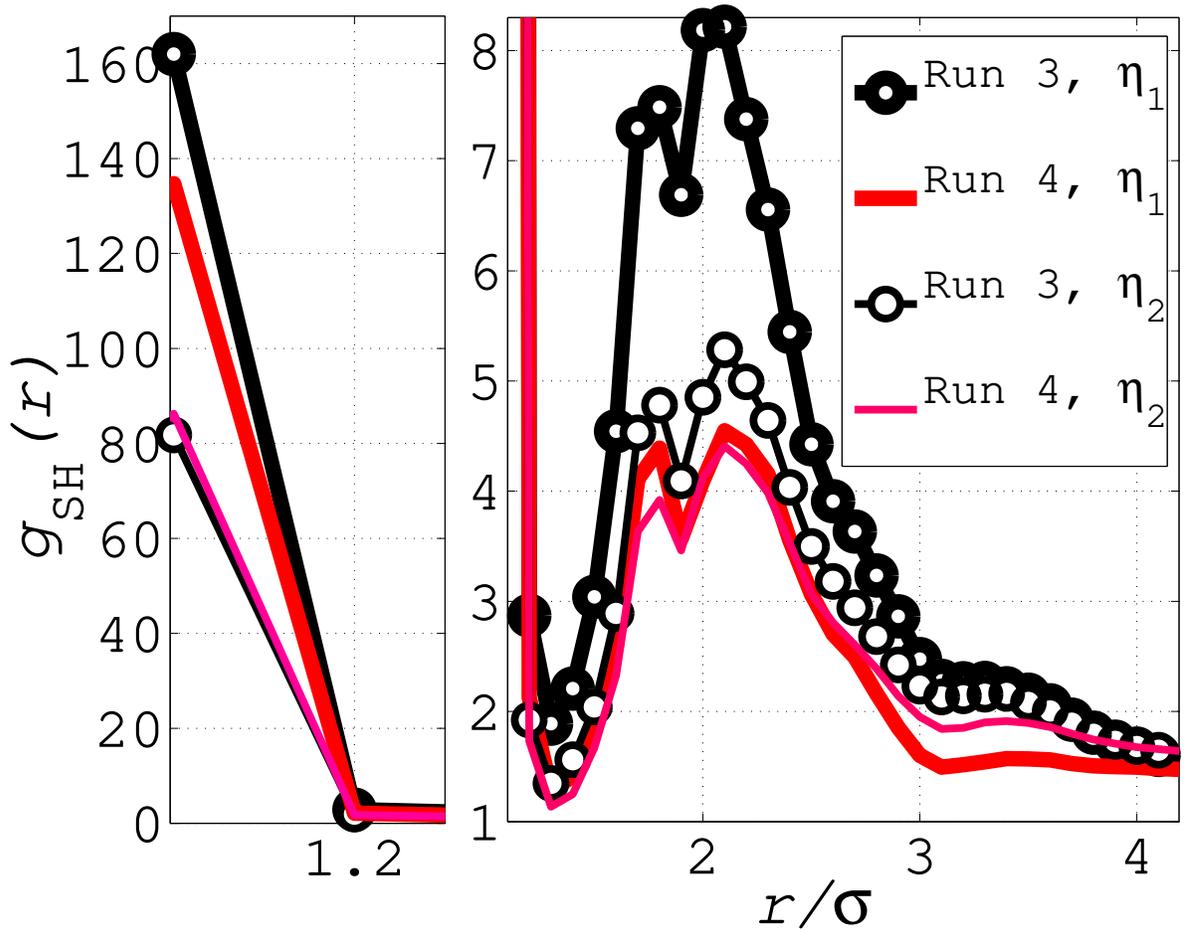}
\caption{(Color online) Sulfonate-proton correlation
 function   $g_{SH}(r)$ as a function of the separation distance $r$. The
 first and second peak areas are shown separately. Thick lines are for
 the membrane $\eta_1$, thin lines are for the membrane $\eta_2$. Line
 with symbols- Run~3, full lines- Run~4. 
 \label{fig4-gsh}}
\end{figure}

The proton mobility in solvated ionomers is strongly affected by
proton--head group association effects. On the
one hand, this association localizes the protons near the head groups, and
therefore decreases the rate of vehicular diffusion across the membrane. On the
other hand, the localization effect increases the rate of the hopping 
diffusion of protons from sulfonate to sulfonate. This so-called
surface diffusion is believed to be additionally enhanced by the 
water--proton electrostatic interactions and the side-chain thermal fluctuations.    
The strength of the proton--sulfonate association is commonly evaluated in
the terms of the proton distribution around the head-groups.   Our simulation
results for the  sulfur-proton pair correlation function $g_{SH}(r)$ are plotted
 in Figure~\ref{fig4-gsh}. The first proton shell,  seen as a very high
 peak on the left side of Figure~\ref{fig4-gsh}, originates from the attractive
 Coulomb forces  between the  protons and the  SO$_3^{-}$
 groups. 
The second peak of  $g_{SH}(r)$ on the right side of
Figure~\ref{fig4-gsh} arises from the proton shells of 
neighboring sulfonates in the multiplet. 
The condensation effect of protons on the sulfonates is  noticeably stronger in
 the  membrane $\eta_1$ than in the case of the membrane $\eta_2$. 
As a consequence,  the proton  mobility in the membrane $\eta_2$ will
be higher. 

The effect of a proton--sulfonate  association
 also depends on the membrane hydration level $\lambda$
\cite{cui-2008}:
 the association is weak for the hydrated membrane with $\lambda=1$,
 whereas it is  strong for the membrane with $\lambda=5$.   
The position of the minimum of $g_{SH}(r)$ corresponds to the position
of the first maximum of $g_{SS}(r)$ in Figure~\ref{fig3-gss-a}.

\begin{figure}
\includegraphics*[width=1.\textwidth]{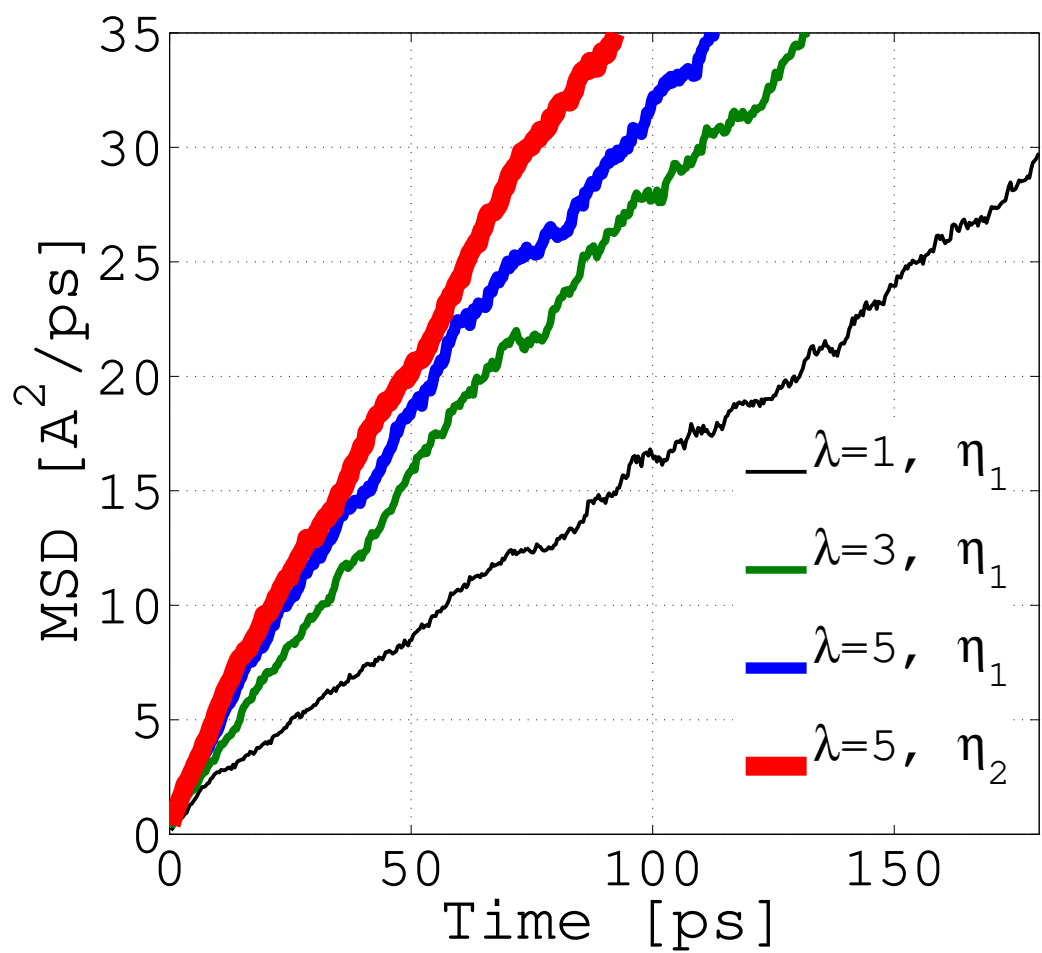}
\caption{(Color online) Mean squared displacement of protons as a function of
 time for Runs~2--4.
 \label{fig12-msd}}
\end{figure}

The mean square displacements (msd) of protons for
different membrane hydrations $\lambda$ are plotted in Figure~\ref{fig12-msd}. 
As  is expected from the proton delocalization effect in hydrated
membranes, higher membrane hydrations result in larger proton
displacements \cite{cui-2007}. The msd result for the membrane $\eta_1$ is
below the corresponding result  for the membrane $\eta_2$ for Run~4. This happens
partly due to the  strong proton delocalization effect, and partly due to the
small nearest-neighbor distances $R_B^S$ in the membrane $\eta_2$.

The calculated values  for the diffusion 
coefficient of protons, 
\begin{equation}
D=  \lim_{t \rightarrow \infty} \frac{\text{msd}(t)}{6t} \,\,\, ,
\label{diffusion}
\end{equation}
 are gathered in Table~\ref{table-eta}. The proton diffusion, similar
 to the proton mobility, is stronger in membrane $\eta_2$ than in
 membrane $\eta_1$ because of the low proton-sulfonate association.  
There are two other factors that contribute to the proton diffusion of
 membrane $\eta_2$: the existence of `bulk'
 sulfonates between neighboring multiplets and  the accumulation of
 water molecules around the multiplets. Both these factors can lead to
 the formation of temporary 
 bridges, sulfonic and/or solvent in nature,  between the multiplets.
 Our results for proton diffusion  are in  good 
 agreement with the simulation results of  Ref.~\cite{cui-2007}.   
However  they are small compared to the proton diffusion coefficients
 experimentally observed in fully hydrated  Nafion ionomers. This
 discrepancy is most probably not due to our neglect of the Grotthuss
 mechanism, which is strongly suppressed when $\lambda$ is small
 \cite{seeliger-2005,eikerling-1997,thompson-2006}, but is a
 consequence of the reduced number of pathways in our low-humidity,
 low-sulfonate model. 

\section{Discussion}
We have investigated the swelling properties of multiplets
in low humidity ionomers with low sulfonate concentration by
examining different models  for the sidechain architecture. Our
primary goal was to determine the dependence of multiplet swelling on
the hydration level $\lambda$ and the sulfonate concentration $\eta$
of the membrane.  

\begin{figure}
\includegraphics*[width=1.\textwidth]{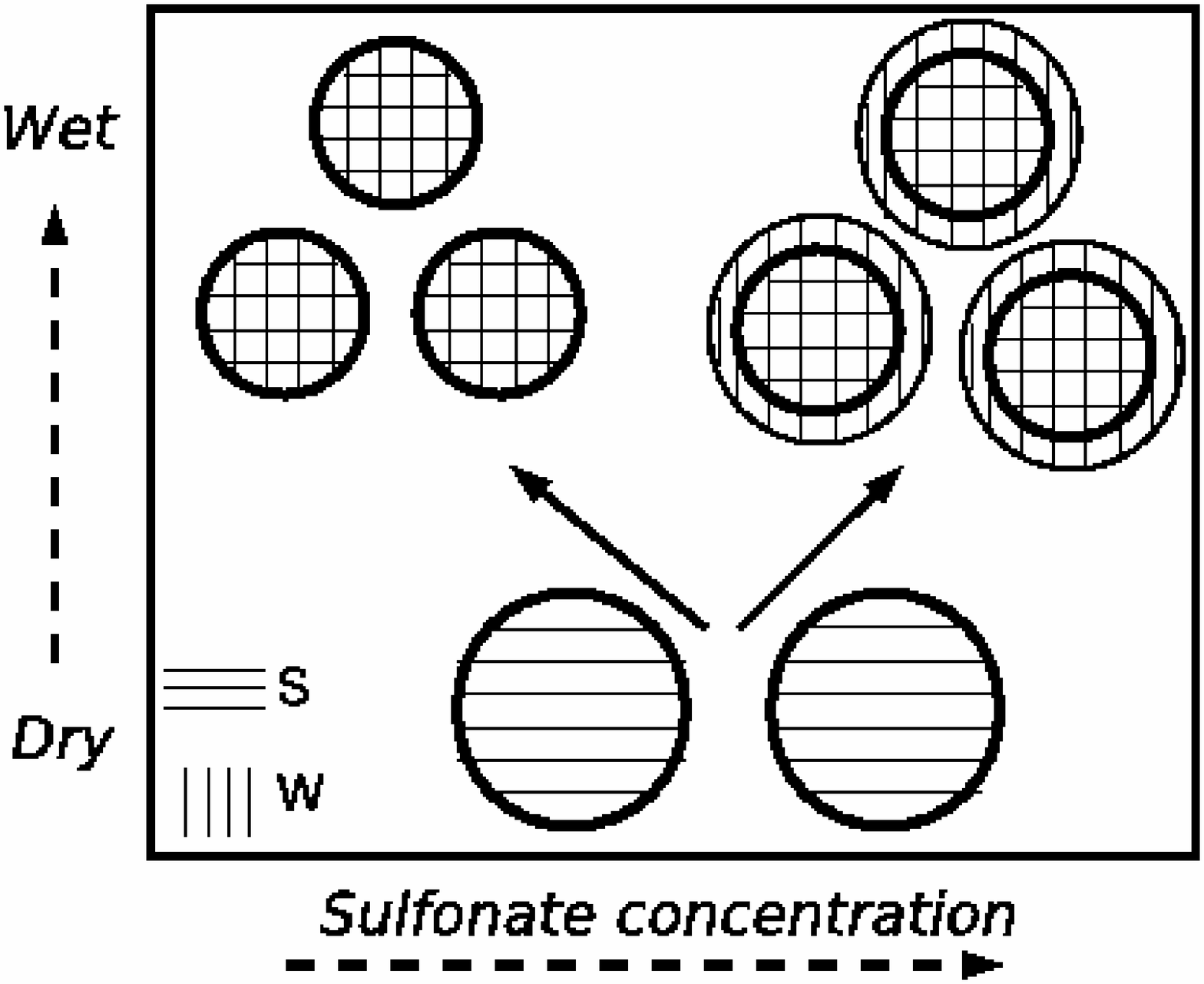}
\caption{ Schematic illustration of the multiplet hydration. At a low
  sulfonate concentration the hydrated multiplets consist of
  sulfonates and water molecules. At a higher sulfonate concentration
  each of the multiplets is surrounded by a water shell. 
The splitting of  dry multiplets into  smaller hydrated multiplets is
also sketched. Vertical/horizontal hatching is used for the
water (W) and the sulfonate (S) areas of the multiplet.
 \label{fig-sketch}}
\end{figure}

Our main result is the fission of the sulfonate multiplets into smaller
parts at the onset of membrane hydration. This behavior
is not explained by the classical theories of cluster swelling in ionomers,
according to which the swelling should be a continuous and monotonic process
of multiplet expansion.
 The resultant small multiplets will consequently swell,  
keeping the number of their sulfonate population constant, when more
water is absorbed into the membrane.  We  have also found that the
location of the of water in low-sulfonate membranes strongly depends
on the sulfonate concentration. For a relatively low sulfonate
concentration nearly all sulfonate groups are in multiplet
formations. The average water loading parameter per multiplet
$\chi_W/\chi_S$, where $\chi_W$ is the number of water molecules
belonging to the multiplet, and $\chi_S$ is 
the number of sulfonates in the multiplet, perfectly matches the
water content of the membrane $\lambda$. However, for  relatively
high sulfonate concentrations, the water loading parameter per
multiplet $\chi_W/\chi_S$ is consistently larger than the parameter
$\lambda$ for the  membrane hydration levels considered.  We assume
that, when the sulfonate concentration approaches the percolation
threshold for head-groups, a fraction of the sulfonates are randomly
distributed between the existing multiplets. These bridging
sulfonates cannot retain their full solvation shell in the hostile
environment of hydrophobic backbones. The excess water molecules
stripped from these `bulk' sulfonates are consequently redistributed
between the existing multiplets. The results of our structural
analysis confirm the formation of unexpected water shells around 
sulfonate multiplets. The multiplet fission and the water
encapsulation effects are illustrated schematically in
Figure~\ref{fig-sketch}. 
      
Our discovery of the  uneven distribution of the water-to-sulfonate
loading  in the ionomer opens a 
new window into the percolation characteristics of the hydrophilic
network in ionomers.  It is no longer sufficient to have a continuous
pathway among sulfonates in order for percolation of protons to occur,
as some of these sulfonates may be found in the hydrophobic material,
where they are not capable of contributing to proton transport.   The
predicted hydration levels necessary for good transport of protons
will thus be higher than they would be if the presence of sulfonates
encapsulated in backbone material were ignored. 

We have also analyzed the structural deformations occurring in the ionomer as a
 result of membrane swelling, and found that in swollen membranes the
 ionomer is in a more relaxed state. The degree of relaxation,
 however, is sensitive to the sulfonate concentration:  the
 sidechains and backbones are found to be  more relaxed in
 low-sulfonate membranes. This result is  a direct consequence of the
 fact that in low-sulfonate membranes the sulfonate cluster is less
 dense, and can relax more readily than in the denser environment of
 the high-sulfonate material. However,  proton diffusion is  stronger
 in high-sulfonate membranes, and can potentially benefit from 
the formation of temporary  solvent and sulfonate  bridges between the
multiplets.

 An interesting question, yet to be resolved, is whether the
 multiplet  splitting and shrinking effects depend on the 
 pretreatment history of the membrane. 
The membrane morphology  is known to be affected by the type
 of pretreatment, such as boiling, annealing, drying, poling,
 stretching,  etc., and by the order in which these steps are taken. In
 most cases the  impact  of the pretreatment is either  the formation of
 a new morphology  with an anisotropy in the backbone and sidechain
 orientations, or  the reshaping of the network of hydrophilic clusters. 
In our current work the dry membrane was `numerically pretreated' by our
fragmentation and de-fragmentation procedures, as described in section
 III. 
We assume that our membrane has a network of hydrophilic pores
resembling the network  in a mold-extruded membrane, provided it has then
been annealed.  

In order to analyze the consequences of the residual anisotropy 
in the ionomer on  the multiplet reorganization effects reported in this work, 
 we carried out test simulations for a poled and dry Nafion-like
 ionomer. According to our previous results on ionomer poling
 \cite{article-2},  rod-like aggregations of head groups are formed
 along the direction of the applied electric field. The poled
 structures were found to  be stable after the release of the poling
 field. One of the poled  
 structures of Ref.~\cite{article-2} was used as a starting
 configuration for Run~1 of our current work.      
 Our simulation result indicated that a similar reorganization effect of
 sulfonate multiplets, as seen in the case of isotropic membranes, takes
 place. Hence, we conclude that the splitting and shrinking effects
 are robust against  structural anisotropy in the membrane.

We also performed test simulations to clarify the nature of
multiplet reorganization in dry membranes that had been previously swollen. 
The hydrated membrane from Run~4 with water content $\lambda$=5 was 
first dried through a simple elimination of  all water molecules in the
simulation box. Then the water-free  membrane was gradually shrunk to
the system size used for Run~1.  The results obtained show that  the initially dry
membrane, membrane I, and the pretreated dry membrane, membrane II, have 
different structures. In the latter the multiplet sizes $R_A^S$
were smaller and close to the multiplet sizes corresponding to Run~2. However, after 
annealing at high temperatures, the discrepancies between the membranes
disappeared, and both membranes exhibited the splitting and shrinking of
multiplets at the onset of hydration.

In future work, we plan to extend the model presented here to take
into account the partial charges on the side-chain monomers. Our preliminary
results indicate that a partial delocalization of the negative charge
along the sidechain head group has a noticeable role in the membrane
swelling process.

\acknowledgments
We thank E. Spohr and R. Wycisk for valuable comments during the preparation of
this paper. This work was supported by the US Department of Energy under grant
DE-FG02-05ER46244 and  by the German Science Foundation (DFG) under grant
LO 418/12-1.  It was made possible by use of facilities at the
Case ITS High Performance Computing Cluster and the Ohio Supercomputing
Center. 

\newpage
\begin{table} [!ht]
\caption{List of key variables.}
\begin{ruledtabular}
\begin{tabular}{lc} 
$\lambda$ & number of water molecules per sulfonate group \\ 
$\sigma$ & monomer diameter \\
$\epsilon$ & dielectric constant of medium \\ 
$Q_p$,$Q_S$,$Q_{O_3}$ & normalized charges of protons, sulfur atoms
 and head group oxygens  \\ 
$\varepsilon_{LJ}$ &  Lennard-Jones interaction parameter between monomers\\
 $k_b$, $k_{\theta}$, $k_{\alpha}$  & stretching, bending and torsion  force constants \\
$r_0$ &  equilibrium bond length for backbone and sidechain  \\
$\theta_0$ & equilibrium bending angle for backbone and sidechain  \\
$k_B$, $T$ & Boltzmann constant and  system temperature \\ 
$n_1$,$n_2$ & number of hydrophobic and hydrophilic monomers per sidechain \\ 
$n_3$ & number of hydrophobic backbone monomers between adjacent sidechains \\
$\eta$ & molar concentration of head groups \\ 
$N_0$ &  Avogadro's number \\ 
$N$, $N_S$& total number of ionomer monomers,  number of sulfonates  \\ 
$L$, $V$& length of simulation box,  volume of simulation box \\ 
$g_{SS}(r)$, $g_{SH}(r)$ &  sulfonate-sulfonate and sulfonate-proton
 pair correlation  functions \\ 
$ g_{WW}(r)$ &  water-water  pair correlation  function \\
$R_A^S$, $R_A^W$ & size of sulfonate-sulfonate multiplets and water clusters \\
$R_B^S$, $R_B^W$ & nearest-neighbor distance between for  sulfonate multiplets
 and  water clusters \\
$\chi_S$, $\chi_W$ & number of head groups and number of water
 molecules in a multiplet \\
$\xi$  & number of multiplets in the simulation box \\
$l_D$ & electrostatic screening length \\
$\overline l$ & average separation distance between sulfonates \\
$S(q), \, q=2\pi/r$ & sulfonate-sulfonate structure factor \\ 
$\overline R$ & correlation length of density-density fluctuations of sulfonates \\
$P(\alpha)$ & Probability distribution of the dihedral angle along the
 polymer segments \\
$D$ & diffusion coefficient of protons \\ 
$\rho$ & 3D density of membrane components 
\end{tabular}
\end{ruledtabular} 
\label{table-1} 
\end{table}

\newpage

\begin{table}[!ht]
\caption{\label{table-2} Parameters used in simulation runs. Here 
  $n_1+n_2$ is the total number  of monomers per sidechain, $\lambda$ is the
  water content per head group. }
\begin{ruledtabular}
\begin{tabular}[t]{lcc}
Runs    &   hydration model        & $n_1+n_2$              \\ 
\colrule 
Run~1   &     dry ionomer with no water, $\lambda$=0  &  10  \\ 
Run~2   &   ionomer with explicit water, $\lambda$=1  &  9   \\ 
Run~3   &   ionomer with explicit water, $\lambda$=3  &  9   \\   
Run~4   &   ionomer with explicit water, $\lambda$=5  &  9   \\
\end{tabular}
\end{ruledtabular}
\end{table}

\newpage

\begin{table}[!ht]
\caption{\label{table-eta} Calculated ionomer parameters for
  membranes with sulfonic molar concentration $\eta_1$ and
  $\eta_2$ (shortly called membranes $\eta_1$ and
  $\eta_2$ in the text. The definitions of the parameters used  
  are given in Table~\ref{table-1}.
}
\begin{ruledtabular}
\begin{tabular}[t]{lccccccccc}
 & & & & membrane $\eta_1$ & & & & &\\
 \colrule 
$\lambda$ & $R_A^S$ & $R_B^S$ & $\overline R$ &  $R_A^W$ & $R_B^W$ &
  $\chi_S$ & $\chi_W$ &  $\chi_W$/$\chi_S$ & $D$($\rm cm^2$/$\rm sec$) \\ 
\colrule 
0 & 5.3 & 9.6 & 10  & --  & --  & 16 & -- &-- & --   \\ 
1 & 4.7 & 7.9 & 7.5 & 4.7 & 7.9 & 10 & 10 & 1 & 2.7$\times 10^{-6}$  \\ 
3 & 4.9 & 8.1 & 7.7 & 4.9 & 8.1 & 10 & 30 & 3 & 4.5$\times 10^{-6}$  \\   
5 & 5.0 & 8.5 & 8.0 & 5.0 & 8.5 & 10 & 50 & 5 & 5.2$\times 10^{-6}$  \\
\end{tabular}
\end{ruledtabular}
\vskip 0.1cm
\begin{ruledtabular}
\begin{tabular}[t]{lccccccccc}
 & & & & membrane $\eta_2$ & & & & & \\
\colrule 
$\lambda$ & $R_A^S$ & $R_B^S$ & $\overline R$ &  $R_A^W$ & $R_B^W$ &
  $\chi_S$ & $\chi_W$ &  $\chi_W$/$\chi_S$ & $D$($\rm cm^2$/$\rm sec$)      \\ 
\colrule  
0 & 5.0 & 9.4 & 9.97& --  & --  & 22 & -- & --  & --                 \\ 
1 & 4.2 & 7.9 & 7.64& 4.8 & 7.7 & 13 & 17 & 1.2 & 3.5$\times 10^{-6}$  \\ 
3 & 4.4 & 8.0 & 8.0 & 5.1 & 8.04& 13 & 62 & 4.5 & 5.2$\times 10^{-6}$  \\   
5 & 4.6 & 8.2 & 8.35& 5.2 & 8.32& 13 & 101& 7.8 & 6.5$\times 10^{-6}$  \\
\end{tabular}
\end{ruledtabular}
\end{table}

\newpage

\clearpage
\newpage

\clearpage
\newpage

\clearpage
\newpage

\clearpage
\newpage

\clearpage
\newpage

\clearpage
\newpage

\clearpage
\newpage

\clearpage
\newpage

\clearpage
\newpage

\clearpage
\newpage

\clearpage
\newpage

\clearpage
\newpage

\clearpage
\newpage

\clearpage
\newpage

\newpage
\clearpage


\newpage

\begin{references}

\bibitem{first-ionomer} 
McAlevy, U.S. Patent 2,405,971 (August 29, 1946).

\bibitem{perry} M. I. Perry, T. F. Fuller, 
 J. Electrochemical Society {\textbf{149}}, S59 (2002).

\bibitem{sacca-2006}
A. Sacca, A. Carbone, R. Pedicini, G. Portale, L. D'Ilaro, A. Longo,
A. Martorana, E. Passalacqua,  J. Membrane Science {\bf 278},  105 (2006).

\bibitem{saito-2004}
M. Saito, K. Hayamizu, T. Okada, 
 Phys. Chem. B, {\bf 109}, 3112  (2005).

\bibitem{kreuer-2001} K. D. Kreuer,  Journal of Membrane Science {\bf
  185}, 29 (2001).

\bibitem{smitha-2005}
B. Smitha, S. Sridnar, A. A. Khan, 
J. Polym Sci: part B:  Polymer. Phys. {\bf 43}, 1538 (2005).

\bibitem{first-nafion} D. J. Connolly, W. F. Gresham, U.S. Patent
3,282,875 DuPont Co. (November 1, 1966); D. A. Hounshell, J. K. Smith
Jr., Science and Corporate Strategy- DuPont R\&D, 1902-1980, Cambridge
University Press, Cambridge, UK (1988). 

\bibitem{banerjee} S. Banerjee, D. E. Curtin, J. Fluorine Chemistry
{\bf 125}, 1211 (2004).

\bibitem{Gierke2} W. Y. Hsu, T. D. Gierke, Macromolecules
  {\textbf{15}}, 101 (1982); {\textit{ibid}}   J. Memb. Sci.
  {\textbf{13}}, 307 (1982); T. D. Gierke, G. E. Munn, F. C. Wilson,
  J. Polym. Sci. Phys. Ed. {\textbf{19}}, 1687 (1981). 

 \bibitem{robertson} M. A. F. Robertson, Ph.D. Thesis, University of
 Calgary (1994). 

\bibitem{jalani2006} N. H. Jalani, Ph.D. Thesis, Worcester Polytechnic
  Institute (2006). 


\bibitem{MOmodel}    K. A. Mauritz, C. J. Hora, A. J. Hopfinger,
  Polym. Prepr. (Am. Chem. Soc. Div. Polym. Chem.), {\textbf {19}},
  324 (1978); K. A. Mauritz, C. E. Rogers, Macromolecules
  {\textbf{18}}, 483 (1985). 

\bibitem{mauritz2004} K. A. Mauritz, R. B. Moore, Chem. Rev. {\bf 104}, 4535 (2004).

\bibitem{yeager} H. L. Yeager, A. Steck, J. Electrochem. Soc.
  {\textbf {128}}, 1880 (1981). 

\bibitem{eisenberg} A. Eisenberg,  Macromolecules {\textbf{3}},  147
  (1970); Macromolecules {\textbf{30}}, 7914 (1997). 

\bibitem{litt1997} M. H. Litt, Polymer Preprints  {\textbf {38}}, 80 (1997). 

\bibitem{Gebel} G. Gebel, O. Diat, Fuel Cells {\textbf{5}}, 261 (2005);
L. Rubatat, O. Diat, G. Gebel, Physical Chemistry and Soft Matter {\textbf{106}}, 1 (2004).

\bibitem{rohr} K. Schmidt-Rohr, Q. Chen,  Nat. Mater. {\textbf{7}}, 75 (2008).

\bibitem{paddison-1998-epsilon} Paddison, S. J.; Reagor, D. W.;
Zawodzinski, T. A. J. Electroanalyt. Chem. {\bf 91},  459 (1998).
\bibitem{slade} S. Slade, S. A. Campbell, T. R. Ralph, F. C. Walsh, 
J. Electrochem. Soc. {\bf 149}, A1556  (2002).


\bibitem{cappadonia-1994}
M. Cappadonia, J. W. Erning, U. Stimming, 
J. Electroanalyt. Chem. {\bf 376}, 189 (1994).

\bibitem{efield-exp}  H.-L. Lin, T. L. Yu, F.-H. Han,
J. Polym. Research {\bf 13}, 379 (2006).


\bibitem{epsilon-nafion} H.-L. Lin, T. L. Yu, C.-H. Huang, T.-L. Lin, 
J. Polymer Science B: Polymer Physics {\bf 43}, 3044 (2005).

\bibitem{arcella-2005} 
V. Arcella, C. Trogila, A. Ghielmi, 
Ind. Eng. Chem. Res. {\bf  44}, 7646 (2005).

\bibitem{yamamoto-2006}
Y. Yamamotoa, M. C. Ferrari, M. Giacinti,
M. G. Baschetti,  M. G. De Angelis, G. C. Sarti,
 Desalination  {\bf 200}, 636 (2006).



\bibitem{fimrite2005} J. Fimrite, H. Struchtrup, N. Djilali,
  J. Electrochemical Society {\textbf{152}},  A1804 (2005).  

\bibitem{laporta-1999} M. Laporta, M. Pegoraro,  L. Zanderighi, 
Phys. Chem. Chem. Phys. {\bf 1}, 4619 (1999).

\bibitem{lue-2009} S. Lue, S. J. Shieh, J. Macromol. Sci.B: Physics
{\bf 48}, 114 (2009). 

\bibitem{spohr2006} S. Dokmaisrijan, E.  Spohr, J. Mol. Liquids  {\textbf{129}}, 92 (2006).

\bibitem{wescott2006} J. T. Wescott, Y. Qi, L. Subramanian,
  T. W. Capehart, J. Chem. Phys. {\textbf{124}}, 134702 (2006). 

\bibitem{vishnyakovDPD} A. Vishnyakov, A. V. Neimark,  Mesoscale
  Simulations of Hydrated Nafion Membranes, AICHE Annual Meeting,  San
  Francisco (2006).  

\bibitem{article-1} E. Allahyarov, P. Taylor, J. Chem. Phys. {\textbf{127}}, 154901 (2007). 

\bibitem{article-2} E. Allahyarov, P. Taylor, Phys. Rev. E {\bf 80}, 020801(R) (2009).

\bibitem{vishnyakov2001}  A. Vishnyakov, A. V. Neimark,
  J. Phys. Chem. B  {\textbf{105}}, 7830 (2001); {\textit{ibid}} 9586
  (2001). 

\bibitem{allahyarov-stretch}  E. Allahyarov, P. Taylor,
    J. Phys. Chem. B {\bf 113}, 610 (2009). 

\bibitem{lu-2008-epsilon} Z. Lu, G. Polizos, D. D. Macdonald, 
E. Manias, J. Electrochem. Soc. {\bf 155}, B163 (2008).

\bibitem{tsampas-2007} C. G. Vayenas, M. N. Tsampas, A. Katsaounis,
Electrochimica Acta {\bf 52},  2244 (2007).

\bibitem{Paddison1} S. J. Paddison, T. A. Zawodzinski, Solid State
  Ionics {\textbf{113-115}}, 333 (1998). 

\bibitem{paddison-review-1} S. J. Paddison,
  Annu. Rev. Mater. Res. {\textbf{33}}, 289 (2003). 

\bibitem{gebel-moore-2000}
G. Gebel, R. B. Moore, Macromolecules {\bf 33}, 4850 (2000).

\bibitem{paddison2} S. J. Paddison, J. A. Elliott,
  J. Phys. Chem. {\textbf{109}}, 7583 (2005). 

\bibitem{water_parameters} K. Chan, Y. W. Tang, I. Szalai, Molecular
  Simulations {\textbf{30}}, 81 (2004).

\bibitem{jang2004} S. S. Jang, V. Molinero, T. Cagin, W. A. Goddard III,
J. Phys. Chem. B, {\textbf{108}}, 3149  (2004).

\bibitem{artificial-step} A temporary fragmentation of ionomer
  molecules into smaller segments in order to speed up its
  equlibration in molecular dynamics simulations is equivalent of
  heating up the ionomer above its glass transition temperature $T_g$
  in experiments.

\bibitem{rivin-vishnyakov} A. Vishnyakov, A. V. Neimark,
  J. Phys. Chem. B {\textbf{104}}, 4471 (2000). 

\bibitem{rivin2004} D. Rivin, G. Meermeier, N. S. Schneider,
  A. Vishnyakov, A. V. Neimark,  J. Phys. Chem. B {\textbf{108}},
  8900 (2004). 

\bibitem{elliott1999} J. A. Elliott, S. Hanna, A. M. S. Elliott,
  G. E. Cooley, Phys. Chem. Chem.  Phys. {\textbf{1}}, 4855 (1999). 

\bibitem{Lekner}  M. Mazars,  J. Chem. Phys. {\textbf{115}}, 2955 (2001). 

\bibitem{cui-2007} S. Cui, J. Liu, M. E. Selvan,
D. J. Keffer, B. J. Edwards, W. V. Steele, 
J. Phys. Chem. B {\bf 111}, 2208 (2007).

\bibitem{cui-2008} S. Cui, J. Liu, M. E. Selvan,
S. J. Paddison, D. J. Keffer, B. J. Edwards, 
J. Phys. Chem. B {\bf 112}, 13273 (2008).

\bibitem{brunelo-2009}
G. Brunello, S. G. Lee, S. S. Jang,  Y. Qi,
J. Renewable and Sustainable Energy {\bf 1}, 033101 (2009).

\bibitem{spohr-2004}
E. Spohr, Mol. Simul. {\bf 30},  107  (2004).

\bibitem{elliott2006} J. A. Elliott, S. Hanna, A. M. S. Elliott,
  G. E. Cooley, Polymer Engineering and Science {\bf 46}, 228  (2006).

\bibitem{young2002}
S. K. Young, S. F. Trevino, N. C. Tan,
 J. Polymer Science: Part B, Polymer Physics {\bf 40},
387 (2002).

\bibitem{choi2005} P. Choi, N. H. Jalani, R. Datta, 
J. Electrochemical Society {\bf 152},  E123 (2005). 

\bibitem{seeliger-2005} D. Seeliger, C. Hartnig, E. Spohr, 
  Electrochim Acta  {\bf 50}, 4234 (2005). 

\bibitem{eikerling-1997} M. Eikerling, A. A. Kornyshev, U. Stimming,
J. Phys. Chem. {\bf 101}, 10807 (1997).

\bibitem{thompson-2006} E. L. Thompson, T. W. Capehart, T. J. Fuller,
  J. Jorne, J. Electrochem. Society {\bf 153},  A2351 (2006).





\end{references}
\end{document}